\documentclass{optica-article}

\journal{opticajournal} % for journals or Optica Open

\articletype{Roadmap}

\usepackage{lineno}
\usepackage[normalem]{ulem}
\usepackage{orcidlink}
\usepackage{rotating}
\usepackage{placeins}
\usepackage[strict]{changepage}
\usepackage{tabularx}
\usepackage{longtable}
\usepackage{makecell}
\usepackage{upgreek}
\usepackage{amsbsy}
\usepackage{bm}
\usepackage{siunitx}
\usepackage{hyperref}

\usepackage[version=4]{mhchem}

\newcommand{\etal}{\emph{et~al.}}

\providecommand{\pnl}[1]{{\textcolor{black}{(#1)}}}%command for panels in figures

\linenumbers % Turn off line numbering for Optica Open preprint submissions.

\nolinenumbers

\begin{document}

\title{Crystalline metal flakes: Platforms for advanced plasmonics and hybrid 2D material architectures}

\author{Sergejs Boroviks\,\orcidlink{0000-0002-3068-0284}\authormark{1}}

\author{Siarhei Zavatski\,\orcidlink{0000-0003-4530-4545}\authormark{1}}

\author{Thorsten Feichtner\,\orcidlink{0000-0002-0605-6481}\authormark{2}}

\author{Jer-Shing Huang\,\orcidlink{0000-0002-7027-3042}\authormark{3}}

\author{Olivier J. F. Martin\,\orcidlink{0000-0002-9574-3119}\authormark{1}}

\author{Bert Hecht\,\orcidlink{0000-0002-4883-8676}\authormark{2}}

\author{N.~Asger~Mortensen\,\orcidlink{0000-0001-7936-6264}\authormark{4,5,*}}

\address{\authormark{1}Nanophotonics and Metrology Laboratory, Swiss Federal Institute of Technology Lausanne (EPFL), EPFL-STI-NAM, Station 11, Lausanne, CH-1015 Switzerland}
\address{\authormark{2}Nano-Optics and Biophotonics Group, Experimentelle Physik 5, Physikalisches Institut, Universität Würzburg, Am Hubland, Würzburg, Germany}
\address{\authormark{3}Leibniz Institute of Photonic Technology, Albert-Einstein-Str. 9, 07745 Jena, Germany}
\address{\authormark{4}POLIMA---Center for Polariton-driven Light--Matter Interactions, University of Southern Denmark, Campusvej 55, DK-5230 Odense M, Denmark} 
\address{\authormark{5}D-IAS---Danish Institute for Advanced Study, University of Southern Denmark, Campusvej 55, DK-5230 Odense M, Denmark} 
\email{\authormark{*}namo@mci.sdu.dk}

\begin{abstract*} 
Crystalline noble metal flakes are emerging as versatile platforms in nanophotonics, enabling a broad range of optical phenomena and applications. Their atomically flat surfaces, high crystallinity, and superior optical quality open new avenues in advanced plasmonics, quantum light generation, and hybrid photonic systems. In contrast to conventional polycrystalline metal films, which typically suffer from higher optical losses due to grain boundaries, surface roughness, and structural disorder, these monocrystalline flakes provide minimal scattering and enhanced performance. They serve as templates for precise nanostructuring through techniques like focused-ion beam (FIB) milling and are crucial for advanced applications in sensing and optoelectronics. Additionally, they facilitate frontier research in quantum plasmonics, enabling fundamental studies of nonlocal optical effects and the generation of nonclassical light. Furthermore, the well-defined $\{111\}$ facets of these flakes host Tamm--Shockley surface states that support 2D plasmons coexisting with bulk modes. At near-infrared wavelengths and beyond, crystalline flakes act as nearly ideal metallic mirrors, featuring surface roughness limited only to atomic terrace steps, making them highly suitable for integration with 2D materials in hybrid photonic architectures. This review surveys the key roles these flakes play, highlighting recent developments and discussing future prospects while emphasizing their unique benefits in addressing fundamental and applied challenges in modern nanophotonics.
\end{abstract*}

%%%%%%%%%%%%%%%%%%%%%%%%%% body  %%%%%%%%%%%%%%%%%%%%%%%%%%

\section{Introduction}

Plasmonic technologies rely on metal nanostructures to manipulate light at the nanoscale~\cite{Lal:2007,Stockman:2011,Fernandez-Dominguez:2017}. Traditionally, polycrystalline -- evaporated or sputtered -- metal films have been employed, driven largely by advancements in nanofabrication technologies~\cite{Henzie:2009} for both the physical vapor deposition (PVD) of metal films on substrates and their subsequent nanopatterning using sophisticated lithographic techniques, including electron-beam lithography (EBL) and focused-ion beam (FIB) milling. Despite inherent limitations in optical performance due to surface roughness, grain boundaries, and material inhomogeneity, these techniques have led to significant developments in nanoplasmonics and plasmon-based metasurfaces. However, the relatively high Ohmic losses of metals~\cite{Johnson:1972}, in contrast to the moderate optical losses of dielectrics and semiconductors, have often been perceived as an Achilles' heel for plasmonics, subject to debate on the damping and how to possibly mitigate it~\cite{Khurgin:2015c,GarciadeAbajo:2015,Ndukaife:2016,Boriskina:2017}.

Crystalline metal flakes -- typically noble metals such as gold (\ce{Au}) and silver (\ce{Ag}), or metals like aluminum (\ce{Al}) or copper (\ce{Cu}) -- offer a high-purity, atomically flat alternative that potentially addresses many of these limitations. This has stimulated efforts in better plasmonic films~\cite{McPeak:2015} and the pursuit of crystalline metal flakes in plasmonics has been intuitively linked to their anticipated lower optical losses. Seeded by the seminal work by Huang \emph{et~al.}~\cite{Huang:2010}, crystalline metal flakes are now increasingly utilized in a range of nanoplasmonic applications where minimizing Ohmic losses is critical, spanning advanced devices and structures for frontier plasmonic experiments~\cite{Schmidt:2012,Aeschlimann:2017,Spektor:2017,Frank:2017,Munkhbat:2021,Wu:2022,Zurak:2024,Qin:2025} to sensing~\cite{Sweedan:2024} and plasmon-enhanced quantum optics~\cite{Bozhevolnyi:2017a,Bozhevolnyi:2017b,Fernandez-Dominguez:2018}.

However, the advantages of crystalline metal flakes extend beyond merely achieving slightly lower Ohmic losses. Flakes with sub-micron lateral dimension can be used as plasmonic cavities with SPPs being in-plane confined by the flake edges~\cite{Viarbitskaya:2013,Frank:2017}, while larger flakes also serve as critical components in fundamental explorations of out-of-plane extremely confined polaritons~\cite{Boroviks:2022,Menabde:2022a}, where surface-roughness-induced scattering is minimized due to the atomic flatness of the flakes. Furthermore, crystalline metal flakes exhibit a rich anisotropic nonlinear optical response that reflects the underlying crystal lattice symmetry~\cite{Boroviks:2021,Boroviks:2025}, a symmetry feature absent in polycrystalline samples that exhibit an isotropic response. Furthermore, the fabrication of complex geometries benefits greatly from the flakes’ crystallinity, as the material’s resistance to etching and milling remains uniform over large areas.

Finally, the long-standing fascination with surface states in surface science~\cite{Inglesfield:1982} -- partly seeded by seminal scanning-tunneling microscopy (STM) of quantum corrals for surface-state electrons~\cite{Crommie:1993} -- is now also showing implications for plasmonics, with Tamm--Shockley surface states~\cite{Tamm:1932,Shockley:1939} being revived in the context of electrodynamics~\cite{Pitarke:2007,Mortensen:2021a}, including both extremely confined 2D plasmons~\cite{Suto:1989,Echenique:2001,RodriguezEcharri:2019} and the general electromagnetic surface response of crystalline $\{111\}$ noble metal surfaces~\cite{RodriguezEcharri:2021a, Uskov:2023,Monticone:2025}. In this context, it is also worth highlighting the connection to crystalline noble-metal nanoparticles, which can exhibit a variety of well-defined facets~\cite{Quan:2013}, giving rise to morphology-dependent plasmonic resonances~\cite{Myroshnychenko:2008a,Yoon:2019}.

\section{Chemical synthesis and crystal morphology}

As this review concentrates on the application of crystalline metals in plasmonics and photonics, we will offer only a brief overview of the chemical methods that facilitate the routine synthesis of \ce{Au} flakes. For more comprehensive information on the synthesis and structural characterization of crystal metal films and flakes, we direct readers to papers and reviews that discuss this topic in greater detail~\cite{Xia:2009, Li:2014, Hu:2015, Hoffmann:2016, Grzelczak:2008, Grzelczak:2020, Neal:2021, Yu:2023a, Scarabelli:2023, Sweedan:2025}.

Although the first reports on synthesis and optical characterization of \ce{Au} nanoparticles date back to the time of Faraday~\cite{Faraday:1857}, a rigorous understanding of their nucleation and growth mechanisms was commenced only in the second half of the 20\emph{th} century. A seminal work by Turkevich \emph{et~al.}~\cite{Turkevich:1951} paved the way for improved control over the nanoparticle morphology, shape, size, surface properties and other properties. 
Overall, the yield and size of the nanoparticles are governed by an intricate interplay between nucleation and diffusional growth. These two competing processes depend on the concentration of the precursor in the growth solution: nucleation dominates at supersaturation, whereas diffusional growth prevails at lower concentrations, which is known as the LaMer mechanism~\cite{Lamer:1950}. Typically, synthesis begins with a supersaturated solution, and the concentration is depleted with time as more and more crystal seeds are formed during the burst nucleation stage. Once the threshold concentration is reached, the nucleation slows down and diffusional growth becomes dominant: the flux rate of adatoms to the existing crystal seeds is higher than the rate of formation of new seeds.

An important milestone in the synthesis of anisotropic \ce{Au} nanoparticles was achieved by Brust and colleagues~\cite{Brust:1994}, who developed a two-phase solution synthesis procedure for producing plate-like \ce{Au} nanoparticles. This method, known as the Brust--Schiffrin procedure, was later refined to yield \ce{Au} flakes with very high aspect ratios and lateral dimensions on the order of hundreds of micrometers~\cite{Radha:2011,Radha:2012}. Later, several alternative polyol-based methods were developed, enabling even the direct growth of flakes on substrates from a single-phase solution, and producing flakes with higher aspect ratios~\cite{Li:2005,Guo:2006,Kan:2006,Wu:2015,Ye:2020}.
The majority of the reported synthesis procedures involves reduction of the precursor, typically aqueous solution of gold salt -- hydrogen tetrachloroaurate (\ce{HAuCl4}), also known as chloroauric acid -- in an appropriate organic environment. The common synthesis process is polyol, which involves reduction of the gold salt dissolved in ethylene glycol at elevated temperatures. In some cases, the polyol process is assisted by additional chemicals, such as stronger reducing agents, surfactants and halide ions.
Temperature of the process is a crucial parameter, which is adjusted such that the reduction lasts on the timescale of hours or even days, which ensures slow but steady crystal growth~\cite{Krauss:2018}. Alternatively, formation of \ce{Au} flakes by a tetraoctylammonium bromide (TOABr)-assisted recrystallization approach was shown recently~\cite{Wu:2024}.

Intriguingly, high purity plate-like \ce{Au} crystal samples were also found in nature. Such flakes were first discovered by Hough \emph{et~al.}, in a quartz vein excavated in the Golden Virgin pit, southern Western Australia, and are assumed to form during weathering of metal-rich minerals~\cite{Hough:2008}.

\begin{figure}[ht!]
    \centering
    \includegraphics[width=.9\linewidth]{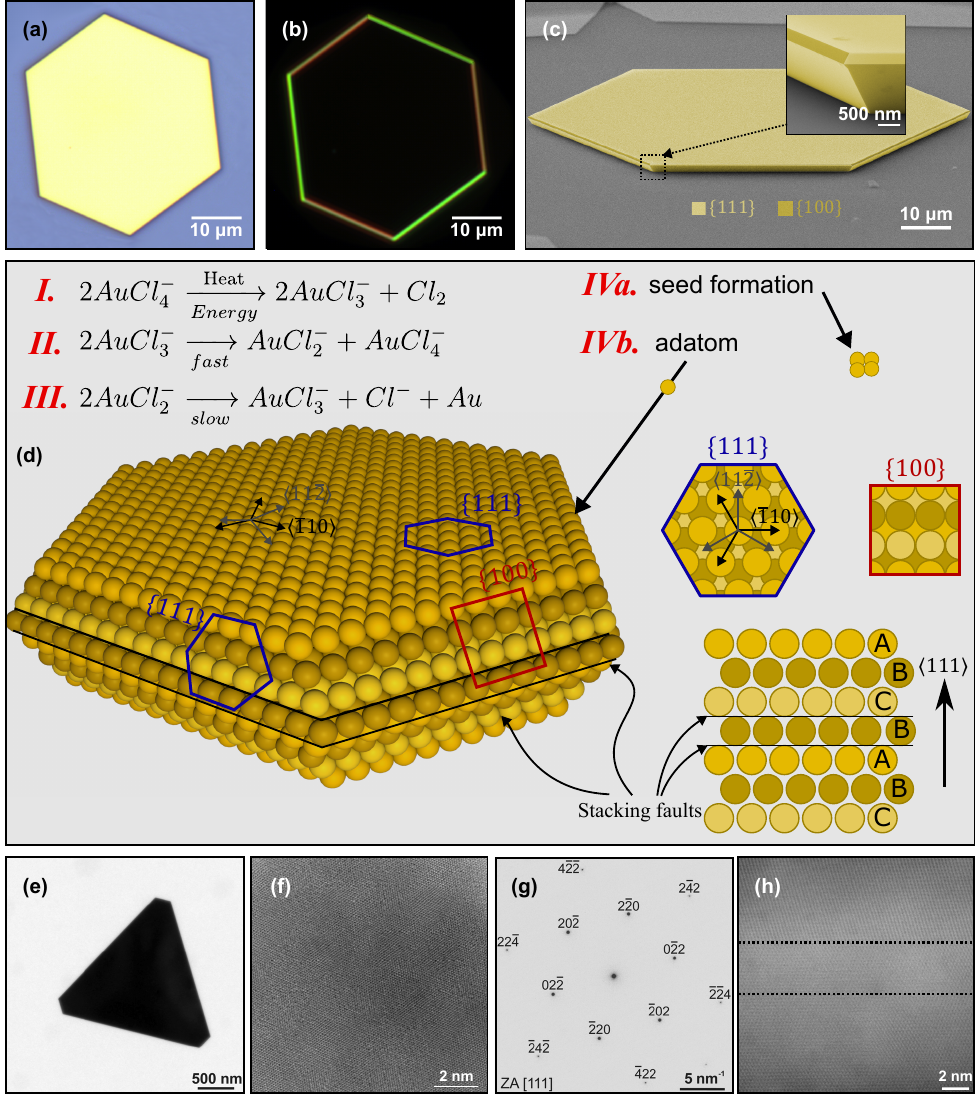}
        \caption{Morphology and crystal structure of \ce{Au} flakes. \pnl{a} Bright-field, \pnl{b} dark-field optical micrographs and \pnl{c} $75$-degree tilted view SEM of a typical \ce{Au} flake sample; the inset in \pnl{c} shows a close-up of the indicated corner of the flake. \pnl{d} A schematic drawing of an \ce{Au} flake morphology, exhibiting a typical distribution of the two lattice planes, $\{100\}$ and $\{111\}$-type, as side facets. Additionally, stacking fault defects in the middle are illustrated. Growth model: I-III Chemical reactions leading to the reduction of \ce{Au} atoms that can form crystal seeds (IVa) or bind to the surface of an already existing crystal (IVb). \pnl{e} TEM image of a small flake sample. \pnl{f} High-resolution TEM image of the flake $\{111\}$-type surface. \pnl{g} Selected area electron diffraction (SAED) pattern of $\{111\}$-type facet. \pnl{h} High-resolution TEM image of the flake cross-section, revealing stacking fault defects, highlighted with black dashed lines. \pnl{a--c}~Reprinted with permission from Ref.~\cite{Boroviks:2018}, \pnl{d} partially reproduced with permission from Refs.~\cite{Krauss:2018,Boroviks:2025}, and \pnl{e--h}~reprinted with permission from Ref.~\cite{Hoffmann:2016}.
    \label{fig:CrystalStructure}}
\end{figure}

Figure~\ref{fig:CrystalStructure} shows optical and scanning-electron microscope (SEM) images of a typical \ce{Au} flake sample, along with the schematics of the crystal structure, schematic model of the crystal growth and additional experimental characterization results. Under normal conditions, \ce{Au} forms a face-centered cubic (FCC) crystal, whose shape is defined by its Wulff construction. $\{111\}$-type facets, with hexagonal arrangement of surface atoms, correspond to the lowest surface energy configuration. In turn, $\{100\}$-type facets correspond the second favorable atomic configuration, hence their surface area tends to be minimized during the crystal growth. Hence, \ce{Au} flakes exhibit large hexagonally-shaped facets that correspond to $\{111\}$-type planes, whereas sidewalls are represented by alternating pairs of $\{111\}$ and $\{100\}$ facets (see pseudo-color SEM images in Fig.~\ref{fig:CrystalStructure}). 
Interestingly, Boroviks \emph{et~al.} discovered that upon visual inspection, the dark-field optical images can be used to distinguish different edge structures associated with different crystallographic planes~\cite{Boroviks:2018}.

The key element for achieving high aspect ratio plate-like shapes is the creation of a crystal seed that has a particular structural defect: a parallel pair (or more) of so-called stacking faults, also referred to as twin planes~\cite{Elechiguerra:2006, Millstone:2009, Lofton:2005, Perala:2013, Personick:2013}, which are schematically illustrated in Fig.~\ref{fig:CrystalStructure}\pnl{d}. These defects correspond to the deviation in the stacking order of the atomic planes along the $\{111\}$-axis direction and essentially break the symmetry of the crystal, allowing anisotropic growth. Therefore, strictly speaking, plate-like particles do not fall into the category of monocrystals (single crystals), but rather render as quasi-monocrystals.
Fig.~\ref{fig:CrystalStructure}\pnl{h} shows a high-resolution transmission-electron microscopy (TEM) image of a flake cross-section revealing the broken stacking of the atomic planes.

The aspect ratio of the plate-like \ce{Au} flakes is a parameter that is particularly hard to tune. In principle, the high aspect ratio samples are achieved via slowing down the growth in $\langle 111 \rangle$-type axes direction, i.e., by selectively covering ${111}$-type facets with capping agents, and accelerating growth in $\langle 100 \rangle$ axes direction.
In this connection, Krauss \emph{et~al.} presented a comprehensive statistical study on the morphology control of \ce{Au} platelets synthesized via wet-chemical methods~\cite{Krauss:2018}. Their work highlighted the critical influence of growth temperature and chemical environment on the relative growth rates of the platelet’s top and side facets. By tuning these parameters, independent control over thickness and lateral dimensions was achieved, producing ultrathin platelets with thicknesses down to 20\,nm and edge lengths approaching 200\,$\upmu$m. Qin~\emph{et~al.} reported improved control over the flake thickness by injecting the growth solution into lamellar bilayer membranes of nonionic surfactant~\cite{Qin:2013}.
More recently, efforts to synthesize large-aspect-ratio flakes (up to $\sim 10^4$) using gap and halide-assisted methods were reported by Kiani~\emph{et~al.}~\cite{Kiani:2022}. 
In another recent example, Pan \emph{et~al.} have developed an atomic-level precision etching technique to overcome the typical trade-off between lateral size and thickness in wet-chemical synthesis of \ce{Au} flakes~\cite{Pan:2024}. Despite apparent increase of surface roughness, as inferred from atomic-force microscopy (AFM), this method enables the fabrication of ultra-thin, large-area (over $10^4$\,$\upmu$m$^2$) monocrystalline \ce{Au} flakes with thicknesses reduced to essentially few-atomic-layer scale, while featuring excellent mechanical properties~\cite{Zhang:2025}. As will be discussed in greater detail in following sections, maintaining crystallinity, morphological integrity and surface smoothness of flakes with such extreme thickness paves the way for clean mesoscopic experiments revealing exotic effects in plasmonics.

Table~\ref{tab:AuChemSynth} summarizes the main advancements in \ce{Au} flake chemical synthesis methods in chronological order.
While far from being an exhaustive presentation of the existing literature, an obvious trajectory of methodological refinement in the synthesis of crystalline metal flakes is clearly visible. 
The evolution from early-stage surfactant-free approaches to precisely tuned growth protocols improves control over lateral size, thickness, facet orientation, and reproducibility of the samples.
Apart from \ce{Au}, which is primarily discussed in this section, similar efforts are being implemented for the synthesis of \ce{Ag}~\cite{Jin:2003,Xu:2007,Pastoriza-Santos:2008,Deckert-Gaudig:2009,Cai:2010,Rycenga:2011,Kelly:2012,Kumar:2012,Chang:2014,Lyutov:2014,Wang:2015,Zhang:2019a,Chen:2024c}, \ce{Al}~\cite{Solti:2024,Dhindsa:2022,Castilla:2022} and \ce{Cu}~\cite{Tang:2017,Kim:2021,Dayi:2025} flakes.

\begin{table}[ht!]
    \centering
    \begin{adjustwidth}{-1.5cm}{-1.5cm}
    {\small
    
\begin{tabular}{|l|l|m{4cm}|m{4cm}|l|l|}
    \hline
    Year & Reference & Method & Feature & \makecell[l]{Typical side\\ length} & \makecell[l]{Typical \\ thickness} \\
    \hline
    2005 & Ah~\etal \cite{Ah:2005} & Reduction with sodium citrate in water & First report of "machinable" flakes & $\sim\SI{310}{\nano\meter}$ & $\sim\SI{28}{\nano\meter}$\\
    2006 & Li~\etal \cite{Li:2006} & Polyol in ethylene glycol using polyvinylpyrrolidone & Size and shape control via concentrations and temperature & $\lesssim\SI{20}{\micro\meter}$ & $\gtrsim\SI{70}{\nano\meter}$\\
    2006 & Guo~\etal \cite{Guo:2006} & Polyol in ethylene glycol at \SI{90}{\degreeCelsius} & No capping agents/surfactants & $\lesssim\SI{15}{\micro\meter}$ & $\sim\SI{50}{\nano\meter}$\\
    2006 & Kan~\etal \cite{Kan:2006} & Polyol in ethylene glycol at \SI{150}{\degreeCelsius} & Demonstrated thermal stability & $\lesssim\SI{50}{\micro\meter}$ & $\sim\SI{70}{\nano\meter}$\\
    2007 & Kawasaki~\etal \cite{Kawasaki:2007} & Thermolysis in two-component ionic liquid at \SI{220}{\degreeCelsius} & First report of millimeter scale-flake & $\lesssim\SI{50}{\micro\meter}$ & $\sim\SI{70}{\nano\meter}$\\
    2010 & \makecell[l]{Radha~\etal \\ \cite{Radha:2010, Radha:2011, Radha:2012}} & Development of the Brust--Schiffrin method~\cite{Brust:1994}, thermolysis at \SI{130}{\degreeCelsius} &	On-substrate growth & $\gtrsim\SI{100}{\micro\meter}$ & $\lesssim\SI{1}{\micro\meter}$\\
    2011 & Huang~\etal \cite{Huang:2011} & Epitaxial growth on graphene oxide film via polyol in hexane/ethanol at \SI{55}{\degreeCelsius} and with 1-amino-9-octadecene surfactant & Synthesis of hexagonal close-packed allotrope crystals & $\sim\SI{500}{\nano\meter}$ & $>\SI{2.4}{\nano\meter}$\\
    2014 & Niu~\etal \cite{Niu:2014} & Synthesis in a layered lamellar hydrogel & Improved control over thickness & $\gtrsim\SI{10}{\micro\meter}$ & $\gtrsim\SI{3.6}{\nano\meter}$\\    
    2015 & Wu~\etal \cite{Wu:2014} & Polyol in ethylene glycol at \SI{50}{\degreeCelsius} & On substrate growth and regrowth & $\lesssim\SI{50}{\micro\meter}$ & $\gtrsim\SI{15}{\nano\meter}$\\       2018 & Krauss~\etal \cite{Krauss:2018} & Polyol in ethylene glycol at varying temperature & Improved control over thickness and lateral size & $\lesssim\SI{180}{\micro\meter}$ & $\gtrsim\SI{15}{\nano\meter}$\\
    2022 & Kiani~\etal \cite{Kiani:2022} & Gap-assisted polyol in ethylene glycol with the help of halides & Improved aspect ratio & $\lesssim\SI{250}{\micro\meter}$ & $\gtrsim\SI{10}{\nano\meter}$\\
    2024 & Wu~\etal \cite{Wu:2024} & Recrystallization of polycrystalline film with tetraoctylammonium bromide & Transformation of deposited films to crystal flakes & $\lesssim\SI{120}{\micro\meter}$ & $\sim\SI{30}{\nano\meter}$\\
    2024 & Pan~\etal \cite{Pan:2024} & Thinning (anisotropic etching using cysteamine solution) of flakes synthesized following method of Krauss~\etal~\cite{Krauss:2018} & Ultrathin flakes, record-high aspect ratio & $\gtrsim\SI{100}{\micro\meter}$ & $\gtrsim\SI{1.9}{\nano\meter}$\\
    \hline
\end{tabular}
}
    \end{adjustwidth}
    \caption{Overview of the pivotal developments in \ce{Au} flake chemical synthesis methods (in chronological order). A more detailed account is provided in the supporting information (Table~\ref{tab:SI}).}
    \label{tab:AuChemSynth}
\end{table}

Finally, we mention a few alternative approaches for the fabrication of high-quality crystalline metal films, such as chemical-vapor deposition (CVD) and atomic-layer deposition (ALD)~\cite{Parkhomenko:2025}, electrochemical deposition~\cite{Grayli:2020,Grayli:2024}, molecular beam epitaxy of \ce{Al} on lattice-matching substrate~\cite{Liu:2015,Chang:2022,Quynh:2022,Raja:2020}
electrochemical etching~\cite{Roy:2010}, various epitaxy methods~\cite{Baski:1994,Silva:2007, Wu:2014, Marconi:2016, Cheng:2019, Rodionov:2019,Abd-El-Fattah:2019, Yakubovsky:2019, Demille:2022, Neal:2022, Grayli:2023, Celebrano:2024}, Czochralski pulling method~\cite{Uecker:2014,Mateck:2013, Jeong:2014}, and even formation of single-atom-thick \ce{Au} sheets -- so-called goldene -- achieved through wet-chemical exfoliation~\cite{Kashiwaya:2024} (although the validity of experimental proof was debated~\cite{Sharma:2026}), and epitaxial growth of ultrathin hexagonal close-packed allotropes~\cite{Huang:2011,Fan:2015}.
Another alternative method for fabrication of high-quality, although polycrystalline, metal films is template stripping, an intriguing technique that has been widely used in surface science to create \ce{Au} films with ultra-large, atomically flat areas suitable for scanning probe microscopy applications~\cite{Vogel:2012}. Early demonstrations achieved surface roughness as low as 2--3\,{\AA} over areas as large as 25\,$\upmu$m$^2$~\cite{Hegner:1993}. Notably, this method has also enabled the production of metal films for plasmonics with state-of-the-art optical losses, even when applied to polycrystalline films~\cite{McPeak:2015}.

\section{Optical and plasmonic properties}
\label{sec:OpticalProperties}
\begin{figure}[ht!]
    \centering
    \includegraphics[width=.9\linewidth]{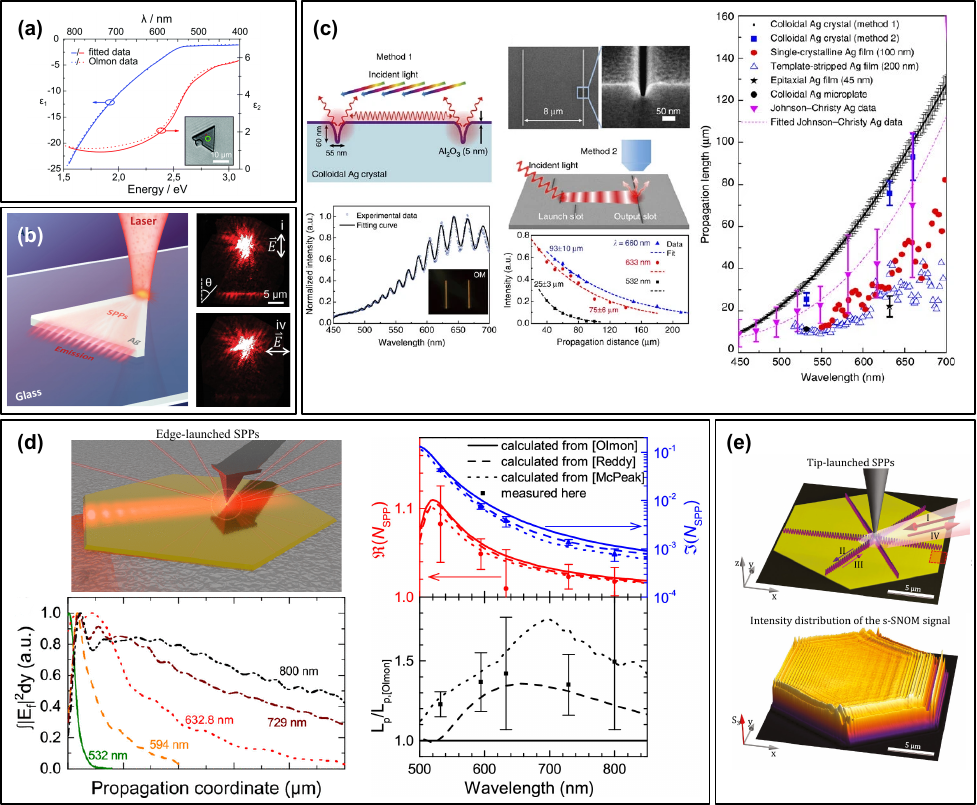}
    \caption{Linear optical properties of crystalline flakes measured using far-field techniques \pnl{a--c} and using SNOM \pnl{d,e}. \pnl{a} Micro-ellipsometry measurements of the complex permittivity of a thin \ce{Au} flake in comparison with measurements on a bulk crystal from Olmon \emph{et~al.}~\cite{Olmon:2012}. \pnl{b} Excitation and scattering of a SPP from the edges of an \ce{Ag} flake. \pnl{c} illustration of the SPP propagation length measurement on a surface of an \ce{Ag} flake with milled in- and out-coupling slits using (left top) white light interference and (left bottom) direct laser excitation and scattering; (right) comparison of these measurements against other experimental reports. SNOM measurements of SPP propagating on an \ce{Au} flake surface in \pnl{d} transmission configuration and in \pnl{e} reflection configuration. 
    \pnl{a} Reprinted with permission from Ref.~\cite{Hoffmann:2016}, \pnl{b} from Ref.~\cite{Zhang:2019a}, \pnl{c} from Ref.~\cite{Wang:2015}, \pnl{d}~from Ref.~\cite{Lebsir:2022}, and \pnl{e} from Ref.~\cite{Kaltenecker:2020}.
    \label{fig:OptChar}}
\end{figure}

Whilst the advantages of the crystalline flakes in terms of structural and mechanical properties are evident from the various characterizations presented in the previous section, the superiority of their optical properties is less immediately apparent, although it is intuitively expected. For example, ellipsometric measurements of the permittivity of \ce{Au} flakes do not show significant differences with mechanically-polished bulk single crystal samples [except for a small decrease in the imaginary part in the visible wavelength range, see Fig.~\ref{fig:OptChar}\pnl{a}]. Intuitively, this is because in such reflection-type spectroscopic measurements, the total field at the metal surface is nearly zero, and hence deeply subwavelength scale surface irregularities do not play an important role. 

However, when considering electromagnetic modes that are confined to the metal surface, such as surface-plasmon polaritons (SPP), one notices a more substantial difference. Short wavelength and strong near-field enhancement makes the SPP, as well as other evanescent modes, extremely sensitive to the surface defects that lead to both radiative and nonradiative losses. 
Figure~\ref{fig:OptChar}\pnl{b,c} show examples of the experiments involving the excitation of SPP modes using different techniques: laser excitation at flake edges~\cite{Zhang:2019a} (similar experiments reported in~\cite{Major:2013,Luo:2015,Kumar:2020a}) or at fabricated gratings/coupling elements~\cite{Wang:2015} (similar experiments reported in~\cite{Boroviks:2019,Dreher:2024}), white light interferometry~\cite{Wang:2015} (similar experiments in~\cite{Wu:2014,Cheng:2019}) and excitation with an electron beam~\cite{Qin:2013}. SPP propagation measured on \ce{Ag} flake surfaces, as shown in Fig.~\ref{fig:OptChar}\pnl{c}, exceeds by on the order of \SI{10}{\micro\meter} the values predicted using the widely employed ellipsometric data of Johnson and Christy~\cite{Johnson:1972}. However, all these examples relied on the SPP out-coupling to the far-field either at the flake edges or via fabricated coupling elements (i.e., gratings, grooves or ridges). This renders quantitative measurements of the SPP propagation length somewhat ambiguous, as the detected far-field intensity depends strongly on the in- and out-coupling efficiencies, which are highly sensitive to the imperfections of the in- and out-coupling elements.

A more direct method for the observation of SPP propagation is scanning near-field optical microscope (SNOM). Figure~\ref{fig:OptChar}\pnl{d} shows SPP propagation characterization using scattering-type SNOM in transmission configuration for five different wavelengths in the visible -- near-infrared bands~\cite{Lebsir:2022}, and Fig.~\ref{fig:OptChar}\pnl{e} in reflection configuration. Transmission measurements are particularly accurate, as they allow the direct recording of both the amplitude and phase of the propagating SPP. In such measurements, the SPP is launched, for example, at the edge of a flake, and detected via the scattering of its evanescent field by a SNOM tip. In contrast, a reflection measurement is somewhat less direct, since in this configuration both excitation and detection of the SPP is mediated by the SNOM tip [see schematics in the upper part of Fig.~\ref{fig:OptChar}\pnl{e}]. Yet another possibility is illumination through a fiber coupled to an aperture tip, which allows point dipole-like excitation~\cite{Abbasirad:2022,Abbasirad:2022a}.
Thus, the SPP wavelength and propagation length may only be deduced from the interference fringes, which are observed in the vicinity of the flake edge due to the superposition of a wave launched by the tip and wave reflected at the edge.
Nevertheless, both of the reports mentioned above, as well as similar works~\cite{Pramassing:2020,Kaltenecker:2021, Casses:2022}, demonstrated a noticeable improvement of the SPP propagation length in comparison with the measurements on polycrystalline samples, overall fitting well with the prediction from ellipsometric measurements of a monocrystalline sample by Olmon~\cite{Olmon:2012}, as can be seen from Fig.~\ref{fig:OptChar}\pnl{d}.

While this section focused on SPPs propagating along metal-dielectric interfaces, the next section will address (partially) localized surface plasmon (LSP) modes. In particular, we will review several examples of nanostructures fabricated from crystalline \ce{Au} that exploit its superior plasmonic performance. Because LSP modes exhibit even stronger near-field confinement than propagating ones (with localization occurring only perpendicular to the direction of propagation), they are correspondingly more sensitive to surface roughness and structural imperfections of the material, making the use of crystalline flakes especially advantageous. 
Incidentally, even flake samples that appear defective can be valuable for investigating localized surface plasmon modes; for example, crystals containing hollow defects may support distinctive resonant behaviors~\cite{Chen:2023a}.

In conclusion of this section, we would like to point out that despite the widespread expectation that crystalline flakes exhibit markedly reduced damping, the experimental results reviewed in this section point to more modest improvements overall. Turning from amorphous films to crystalline flakes, a reduction in losses can indeed be observed, but it is in all fairness less pronounced than the initial narrative may imply. As we will discuss below, the principal advantages of monocrystalline structures may lie equally in their improved structural definition and reproducibility, rather than in a substantial decrease in optical losses alone.

\section{Nano and microfabrication with crystalline metal flakes}
\label{sec:NanoMicroFab}
Following chemical synthesis, a high density of diverse flakes is typically grown directly onto a substrate surface, with post-synthesis cleaning performed using standard solvent rinses. Optical microscopy is then used to visually identify and select sufficiently large, defect-free flakes for further processing, enabling the fabrication of high-quality metallic nanostructures. 
The well-defined crystallinity of monocrystalline \ce{Au} enables precise top-down fabrication of high-definition nanostructures. This is especially critical for techniques such as FIB milling or plasma etching, where different crystal orientations exhibit varying resistance to the ion beam. Consequently, nanostructures patterned into polycrystalline \ce{Au} with randomly oriented grains often display unpredictable geometric imperfections. Using chemically synthesized monocrystalline \ce{Au} flakes for FIB-based nanofabrication overcomes these limitations, allowing the creation of structures with fine features over extended areas—for example, plasmonic two-wire transmission lines spanning several micrometers~\cite{Dai:2014a}. 

To the best of our knowledge, the first attempts to pattern \ce{Au} flakes were reported by Ah \emph{et~al.} in 2005~\cite{Ah:2005}, although earlier works may exist.
Since then, significant advances in nanofabrication techniques and high-resolution lithography have made flake nanopatterning far more versatile and precise. 
Alternative nanopatterning methods were also developed in parallel. For example, Wiley \emph{et~al.} used a nanoskiving technique, which involves slicing with an ultramicrotome (diamond knife)~\cite{Wiley:2008}, to produce \ce{Au} nanowires embedded in epoxy. We also mention the possibility for controlled nanoscale reshaping of ultrathin \ce{Au} nanoprisms using a tightly focused femtosecond near-infrared laser~\cite{Viarbitskaya:2015}. 
In this review, we focus on three main approaches: direct patterning via FIB milling, patterning through a mask defined by EBL, and various micro-manipulation strategies, including transfer to different substrates, stacking for heterostructure fabrication, and deposition of other colloidal particles.

\subsection{Focused-ion beam milling}

\begin{figure}[t!]
    \centering
    \includegraphics[width=.95\linewidth]{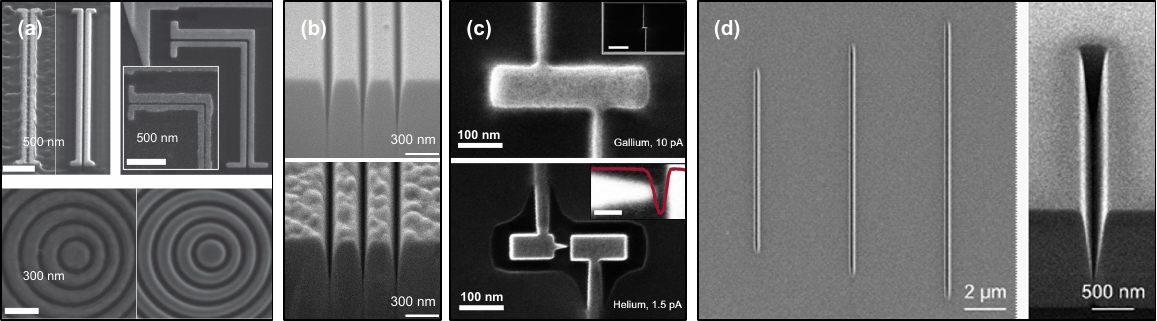}
    \caption{Comparison of \ce{Ga} and \ce{He} FIB milling-based fabrication in poly- and monocrystalline \ce{Au} nanostructures. \pnl{a} (top) Plasmonic waveguides and (bottom) bull's eye resonator and \pnl{b} V-grooves fabricated with \ce{Ga} ion FIB. \pnl{c} \ce{He} ion milling of fine features (asymmetric gap antenna) in a blank predefined with \ce{Ga} FIB. \pnl{d} \ce{He} ion-polished V-groove plasmonic waveguides premilled with \ce{Ga} FIB.
    \pnl{a} Reprinted with permission from Ref.~\cite{Huang:2010}, \pnl{b} from Ref.~\cite{Boroviks:2025}, \pnl{c} from Ref.~\cite{Meier:2023}, and \pnl{d} from Ref.~\cite{Kumar:2020}.
    \label{fig:FIB}}
\end{figure}

Once flakes with suitable geometrical properties -- such as lateral size and thickness -- are identified and localized on the substrate, FIB milling offers a precise method -- down to the single-nanometer scale -- for subsequent patterning~\cite{Hoflich:2023}. The fabrication strategy for high-definition plasmonic nanostructures using large, thin, chemically grown monocrystalline \ce{Au} flakes was seeded by Huang \emph{et~al.}~\cite{Huang:2010}. Unlike PVD-deposited \ce{Au} films (traditionally via sputtering or evaporation), which often exhibit structural imperfections due to their multi-crystalline nature, monocrystalline \ce{Au} flakes allow for the fabrication of ultrasmooth, precisely defined nanoscale features using FIB milling. The difference in fabrication quality is clearly visible in the SEM images shown in Fig.~\ref{fig:FIB}, which compares nominally identical nanostructures in polycrystalline and crystalline \ce{Au} samples. In presented cases, gallium (\ce{Ga}) ion FIB is used to mill various prototype plasmonic nanostructures -- including two-wire transmission lines [Fig.~\ref{fig:FIB}\pnl{a}], asymmetric bull’s eye resonators [Fig.~\ref{fig:FIB}\pnl{a}], meander-shaped nanoantennas [Fig.~\ref{fig:FIB}\pnl{b}], nanoantennas [Fig.~\ref{fig:FIB}\pnl{b}], and V-groove gratings [Fig.~\ref{fig:FIB}\pnl{c}]. 
Visibly, crystalline samples exhibit sharper features, in particular corners and side-walls. Importantly, the FIB milling rate is less homogeneous throughout the polycrystalline samples, which leads to poor control over the milling depth as well as uneven removal of the material causing \ce{Au} residuals around the structures [\emph{gold crumbs}, particularly noticeable in the upper panel of Fig.~\ref{fig:FIB}\pnl{b}].
More recently, with the widespread adoption of noble-gas FIB sources, two-step fabrication procedures have been developed. In these methods, \ce{Ga} FIB is used for coarse structuring in the first step, followed by helium (\ce{He}) ion milling for finer detailing and/or surface polishing. One of the first works applying such a two-step approach to crystalline flakes was reported by Kumar \emph{et~al.}, who fabricated \ce{He}-ion–polished V-groove waveguides in \ce{Au} flakes, as shown in Fig.~\ref{fig:FIB}\pnl{d}~\cite{Kumar:2020}.
In another study, shown in Fig.~\ref{fig:FIB}\pnl{c}, Meier \emph{et~al.} leveraged the precision of \ce{He} ion milling to realize deliberate asymmetries in a nanoantenna blank predefined using \ce{Ga} FIB~\cite{Meier:2023}. The full potential of the combined \ce{Ga} and \ce{He} milling approach is demonstrated in a comprehensive work by Deinhart \emph{et~al.}, which also provides an open-source Python toolbox for automated milling pattern generation and optimization~\cite{Deinhart:2021}.

Kej\'{i}k \emph{et~al.} fabricated and studied arrays of \ce{Au} nanorods on silicon nitride (\ce{Si3N4}) membranes, comparing resonators patterned by FIB milling in both monocrystalline and conventional polycrystalline \ce{Au} films~\cite{Kejik:2020}. In this study, no measurable difference was observed in the $Q$-factor of resonators between polycrystalline and monocrystalline \ce{Au}, which appears to contradict other reports~\cite{Mejard:2017,Liu:2022,Pan:2024}.
However, the TEM images suggest that the FIB milling quality could be further optimized, as some irregularities in the nanorod shapes are observed in both sample types. Additionally, the lack of a significant difference in optical response may be attributed to potential deterioration of the crystalline \ce{Au} due to contamination from \ce{Ga} ions during FIB processing. As a another opposing example, Geisler \emph{et~al.} investigated plasmon transmission along single-crystal gold nanowires and demonstrated quantitative agreement between experimental and simulated decay lengths when all relevant propagation channels were considered. By transferring the fabricated nanostructures onto pristine glass substrates, they were able to exclude additional damping contributions and achieve plasmon propagation lengths approaching simulation predictions~\cite{Geisler:2017}.

Overall, since the early work of Ah \emph{et~al.}~\cite{Ah:2005}, FIB milling has become a widely used technique for fabricating nanostructures from crystalline \ce{Au} flakes. Apart from the aforementioned demonstration, a non-exhaustive list of examples includes various types of gratings~\cite{Lyutov:2014, See:2017, Lin:2019,Ouyang:2021,Sweedan:2024}, advanced plasmonic nanoantennas~\cite{Vesseur:2008,Chen:2014,Celebrano:2015,Klaer:2016,Feichtner:2017,Kullock:2020,Meier:2023,Zurak:2024}, metasurfaces~\cite{Habibullah:2022}, structures for the plasmon excitation~\cite{Frank:2017, Spektor:2017, Davis:2020, Boroviks:2022}, and even micro-drones~\cite{Wu:2022} and micro-robots~\cite{Qin:2025}.
Among the examples mentioned, works that employ a combination of \ce{He} and \ce{Ga} FIB milling stand out~\cite{Chen:2018,Wu:2022,Zurak:2024,Qin:2025,Schurr:2025,Qin:2026,Qin:2026a}. This two-step method is particularly promising, as it has proven to be a reliable approach for obtaining nanostructures of exceptional quality and unprecedented tolerances. As a result, key plasmonic and optical figures of merit (e.g., $Q$-factor or propagation length) can approach values predicted by simulations -- something that is rarely attainable with conventional polycrystalline structures.
Lastly, we note that a general review of the FIB nanofabrication technique for various types of nanophotonic structures can be found in a recent publication~\cite{Yang:2025} (see section 5 therein).

\subsection{Electron-beam lithography}

In an alternative method for the fabrication of crystalline \ce{Au} nanostructures, a pattern is first defined in a resist using EBL and subsequently transferred to the \ce{Au} flake via wet or dry etching. Although FIB offers a more direct, "what you see is what you get" approach, an important advantage of EBL is the absence of \ce{Ga} ion implantation and contamination, such that the intrinsic FCC crystal structure of \ce{Au} is better preserved; in contrast, FIB processing can induce lattice distortions and defects~\cite{Hofmann:2017}.

\begin{figure}[ht!]
    \centering
    \includegraphics[width=.9\linewidth]{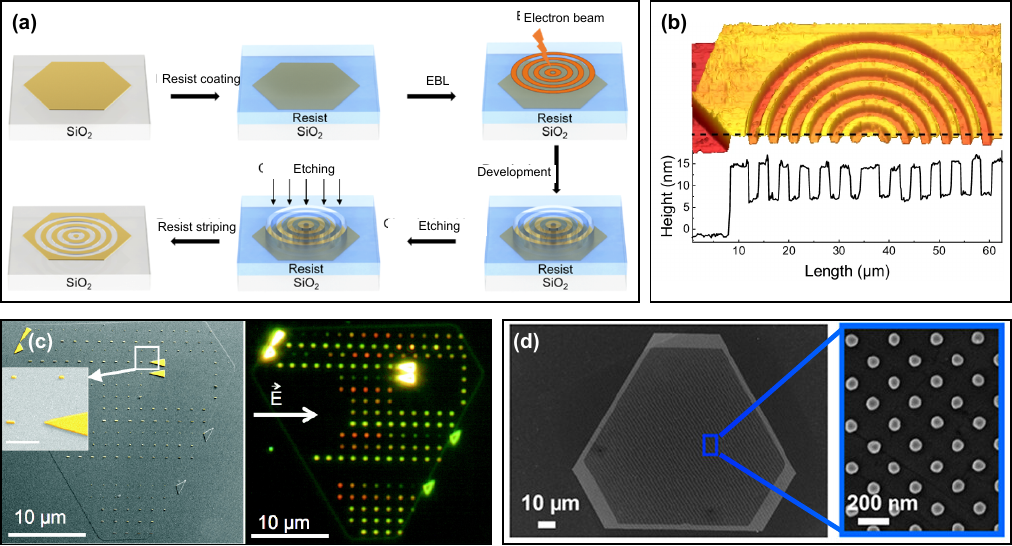}
    \caption{Examples of EBL-based fabrication of crystalline nanostructures. \pnl{a} Typical fabrication process flow for EBL-based nanofabrication with crystalline \ce{Au} flakes. \pnl{b}~AFM image of an \ce{Au} flake patterned with a concentric ring shape using method shown in \pnl{a} and cysteamine as an etchant; lower part of the panel shows a profile along the indicated dashed. \pnl{c} SEM and \pnl{d} dark-field optical microscope images of crystalline \ce{Au} nanoantenna arrays fabricated using EBL and RIE. \pnl{e} SEM image of a crystalline \ce{Au} nanodisk array fabricated using EBL and IBE. \pnl{a,b} Reprinted with permission from Ref.~\cite{Pan:2024}, \pnl{c,d} from Ref.~\cite{Mejard:2017}, and \pnl{e,f} from Ref.~\cite{Kiani:2023}.
    \label{fig:EBL}}
\end{figure}

Figure~\ref{fig:EBL}\pnl{a} shows a typical fabrication process flow, which involves resist spin-coating, pattern exposure with an electron beam, etching and resist stripping. The anisotropic etching can be performed using both dry and wet methods. Two widespread dry etching methods are ion-beam etching (IBE) with argon (\ce{Ar}) ions and reactive ion etching (RIE) with chlorine (\ce{Cl}) or Ar as working gasses. As for the wet enchants, typical reactants are aqueous solution of aqua regia (a mixture of nitric and hydrochloric acid) or iodine-iodide solution~\cite{Green:2022}. Recently, the organic compound-based enchant (cysteamine solution in chloroform) was shown to perform particularly well due to its high selectivity~\cite{Pan:2024}. As mentioned previously, it was used to thin \ce{Au} flakes, and example of a concentric ring nanopattern etched using this method is shown in Fig.~\ref{fig:EBL}\pnl{b}.

As opposed to the widespread lift-off technique, dry etching-based patterning was shown to produces nanostructures with sharper features and better-defined gaps even in polycrystalline films~\cite{Abasahl:2021}. 
In the case of crystalline \ce{Au}, the benefits of dry etching are even more evident: RIE can be facet-selective and the resulting structures have smooth sidewalls~\cite{Greenwood:2022}. For example, M{\'e}jard \emph{et~al.} demonstrated using dark-field spectroscopy that nanoantennas fabricated in crystalline \ce{Au} flakes using RIE [shown in Fig.~\ref{fig:EBL}\pnl{c,d}] exhibit narrower resonances as compared with their polycrystalline counterparts~\cite{Mejard:2017}. In this work, the fabrication process involved an additional step of hard metal mask fabrication using lift-off, since authors employed positive tone resist, specifically polymethyl methacrylate (PMMA) that has poor selectivity in RIE.
More recently, Kiani \emph{et~al.} utilized a negative tone resist, hydrogen silsesquioxane (HSQ), to create a robust mask for \ce{Ar} IBE and fabricate an array of crystalline nanodisks for plasmonic photocatalysis applications~\cite{Kiani:2023} [see Fig.~\ref{fig:EBL}\pnl{e}].

As an alternative to nanopatterning \ce{Au} flakes themselves, the nanostructures can also be created atop of them to exploit their atomically flat surfaces as substrates for plasmonic applications. For example, Siampour~\emph{et~al.} demonstrated dielectric-loaded (DL) SPP waveguides comprised of HSQ with embedded germanium (\ce{Ge}) nanodiamonds with color centers, that worked as quantum emitters~\cite{Siampour:2020}. Waveguides were terminated with designer nanostructures, such as excitation gratings and Bragg mirrors, which allowed to efficiently excite color centers and collect the quantum emission via coupling to DL-SPP modes.
In another example, Boroviks \emph{et~al.} demonstrated gap plasmon-based metasurfaces with an \ce{Au} flake acting as a back-reflector~\cite{Boroviks:2019}. The gap plasmon resonators were defined by lift-off technique in an \ce{Au} layer evaporated atop of a crystalline \ce{Au} substrate covered with a thin silicon dioxide (\ce{SiO2}) spacing layer. Here, substitution of the polycrystalline \ce{Au} substrate with a monocrystalline counterpart resulted in a modest yet measurable (approximately 5\%) efficiency improvement in the visible spectrum, thereby offering a promising route to mitigate plasmonic losses at shorter wavelengths. 

\subsection{Micromanipulation}

\begin{figure}[ht!]
    \centering
    \includegraphics[width=.95\linewidth]{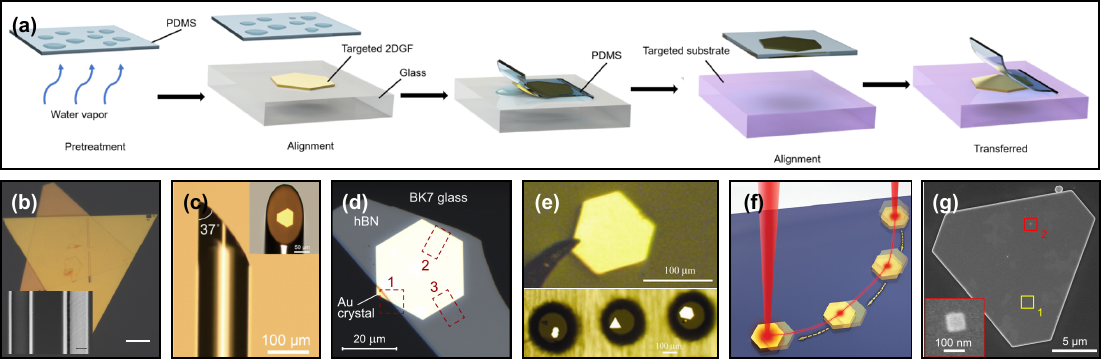}
    \caption{Examples of crystalline flake micromanipulation. \pnl{a} Schematic diagram of \ce{Au} flake transfer from a glass onto another substrate using PDMS stamp. \pnl{b}~Planar gap plasmon waveguide comprised of \ce{Au} flake transferred atop of another flake with a dielectric spacer; inset shows a close-up SEM image of the coupling structure FIB-milled in the lower flake. \pnl{c} Integration of \ce{Au} flake with optical fiber for a background-free excitation of plasmonic nanocavities by the evanescent fields: inset shows an \ce{Au} flake transferred on the core of an angled fibered tip. \pnl{d} Exfoliated hBN flake placed atop of an \ce{Au} flake. \pnl{e} Manipulation of \ce{Au} flake using a pin (top) and \ce{Au} flakes placed in micro-wells for fluorescence enhancement in live cell microscopy. \pnl{f}~Illustration of an omnidirectional translation of \ce{Au} flakes on a solid substrate using a pulsed laser, based on opto-thermo-elastic actuation method. \pnl{g} \ce{Ag} nanocubes spin-coated atop of an \ce{Ag} flake. \pnl{a} Reprinted with permission from Ref.~\cite{Pan:2024}, \pnl{b}~from Ref.~\cite{Boroviks:2022}, \pnl{c} from Ref.~\cite{Liu:2022}, \pnl{d} from Ref.~\cite{Menabde:2022c}, \pnl{e} from Ref.~\cite{Radha:2010}, \pnl{f} from Ref.~\cite{Jia:2024}, and \pnl{g} from Ref.~\cite{Chen:2024c}.
    \label{fig:Transfer}}
\end{figure}

Another important aspect of nano- and microfabrication with crystalline flakes is the ability to deterministically place them on the target substrate and, optionally, interface them with other materials. Fig.~\ref{fig:Transfer} shows several examples of such micromanipulations. 
    
Perhaps the most versatile method -- polymer stamp-mediated transfer -- leverages advancements in handling and transfer techniques originally developed for two-dimensional atomic-layer materials, such as graphene or transition metal dichalcogenide (TMD) materials. These methods now allow for the routine pickup of atomically thin layers from substrates, enabling precise control over their orientation relative to the substrate's normal and facilitating transfer to different surfaces~\cite{Cheliotis:2024}. 
Originally, this method was suggested by Jiao \emph{et~al.} for the transfer of semiconductor nanowires~\cite{Jiao:2008}, however it was proven to be applicable for a wide range of materials and substrates.

Typically, PMMA or polydimethylsiloxane (PDMS) organic polymers are used as stamps. PMMA films are usually employed as \emph{sacrificial} films (i.e., they get dissolved after the transfer of structures to the target sample), whereas PDMS films may be re-usable, as their adhesion to nanostructure can be controlled via temperature or humidity. 
PDMS-mediated transfer of flakes is illustrated in Fig.~\ref{fig:Transfer}\pnl{a}: starting with a substrate sparsely covered with non-overlapping flakes, a water vapor-wetted PDMS stamp is used to pick-up the targeted flake and move it to the target substrate.
Pan \emph{et~al.} demonstrated that this method can be used equally successfully for transfer of plasmonic nanostructures fabricated out of crystalline flakes~\cite{Pan:2024}.
Boroviks \emph{et~al.} used this technique to make heterostructures of \ce{Au} flakes and aluminum oxide (\ce{Al2O3}) films, consisting of two parallel \ce{Au} flakes separated by a few nanometers of \ce{Al2O3} coated using ALD~\cite{Boroviks:2022}, as shown in Fig.~\ref{fig:Transfer}\pnl{b}.
The upper \ce{Au} flakes were transferred from their original synthesis substrates using a commercially available 2D material transfer system developed for the manipulation of 2D materials (from HQ+ graphene). PDMS stamp (WF X4 Gel-Film from Gel-Pak) served as carrier substrates during this process. The transfer was conducted at an elevated temperature of 130$^\circ$C to enhance the adhesion of the \ce{Au} flakes to the target substrate. Liu \emph{et~al.} showed that this method is applicable even in the case of target substrates with very irregular shapes, such as optical fibers [see Fig.~\ref{fig:Transfer}\pnl{c}]~\cite{Liu:2022}. As mentioned above, the polymer stamp method is widely used for the manipulation of two-dimensional materials, and here we mention as an example, interfacing of an \ce{Au} crystal with an \ce{hBN} flake reported by Menabde \emph{et~al.} [see Fig.~\ref{fig:Transfer}\pnl{d}]. More examples of such heterostructures comprised of various two-dimensional materials and crystalline metal flakes are discussed in Sec.~\ref{sec:IdealMirrors}.

A more direct micromanipulation method, however with a high risk of flake damage, is shown in Fig.~\ref{fig:Transfer}\pnl{e}: a micro manipulator-mounted metal needle is used to drag flakes across the substrate and drop them in target positions.
Alternatively, nondestructive and versatile all-optical methods were proposed for manipulation of relatively small flakes (up to approximately~\SI{10}{\micro\meter} in lateral dimension) directly on a solid substrate (e.g. glass or silicon) facing air.
The first demonstration of such on-substrate optical micromanipulation was reported by Lu~\emph{et~al.}, who used a powerful super-continuum light source coupled to a single-mode tapered fiber to displace the \ce{Au} flake samples~\cite{Lu:2017a}. The principle behind this technique lies in balancing between photophoretic forces (due to the light-induced heating) and conventional optical forces (due to the momentum of photons), which allows pushing or pulling the flake in the desired direction. More recently, a simpler method was demonstrated using an opto-thermo-elastic-based actuation, which is achieved using a direct illumination with a focused laser beam in the off-center sample region~\cite{Jia:2024}. As illustrated in Fig.~\ref{fig:Transfer}\pnl{f}, directionality of the actuation is derived from off-centroid illumination by focused laser beams, and the underlying mechanism is due to an interplay between surface friction and opto-thermo-elastic deformations induced by a pulsed laser.
Furthermore, \ce{Au} flake heterostructures with controlled gap sizes can be created in a liquid phase due to Casimir forces between the flakes~\cite{Munkhbat:2021,Schmidt:2023, Kucukoz:2024, Hoskova:2025}, and their position can be controlled using a laser.

The nanoparticle-on-mirror configuration, which has attracted a lot of attention recently due to its ability to sustain extreme optical confinement and enhancement electromagnetic fields~\cite{Leveque:2006,Baumberg:2019}, also requires substrate of high quality and smoothness for robust and reproducible fabrication. Figure~\ref{fig:Transfer}\pnl{g} shows an example of \ce{Ag} nanocubes placed atop of \ce{Ag} flake, which was further used for nonlinear optics experiments~\cite{Chen:2024c}.
Also, Liu \emph{et~al.}~\cite{Liu:2022} and Wang \emph{et~al.}~\cite{Wang:2023a} explored the advantages of using \ce{Au} flakes as mirrors for such plasmonic gap nanocavities, showing that employing atomically smooth \ce{Au} flakes instead of traditional granular polycrystalline films, allows to achieve significant improvements in the optical $Q$-factor and overall performance of the nanocavities (see detailed discussion in Sec.~\ref{sec:PlasmonicNanoantennas}).

\section{Devices and applications}
\label{Sec:devices}

Monocrystalline metal flakes have been used as essential building blocks for high-performance plasmonic devices. The range of applications is not limited to plasmonics and nanooptics, but spans many other fields, including micro- and nano-electronics, mechanics, catalysis and others. The diversity is illustrated in Figs.~\ref{fig:Applications_ElecMech}--\ref{fig:Applications_Quantum}.
In this section we review selected examples of such crystalline plasmonic devices, with primary focus on micro-electromechanical applications, plasmonic nanoantennas and plasmonic circuits, biological and chemical sensing applications, as well as enhancement and coupling to quantum emitters. Applications in nonlinear optics are reviewed in more detail in Sec.~\ref{Sec:nonlinear}.

\subsection{Micro-electromechanical devices}

The high quality of the nanoscale features of these devices were attainable owing to the excellent structural properties and "machinability" of crystalline metal flakes: FIB fabrication of various micromechanical machine elements were demonstrated as early as 2005 by Ah \emph{et~al.}~\cite{Ah:2005} and Yun \emph{et~al.}~\cite{Yun:2005}. Here, we would like to reiterate our remark from Sec.~\ref{sec:NanoMicroFab} that \ce{He} ion-based FIB milling allows to achieve even finer control over single digit nanometer scale features of the plasmonic antennas. In this sub-section we review several examples of such high-quality structures used in micro- and nano-electromechanical devices.

\begin{figure}[ht!]
    \centering
    \includegraphics[width=.9\linewidth]{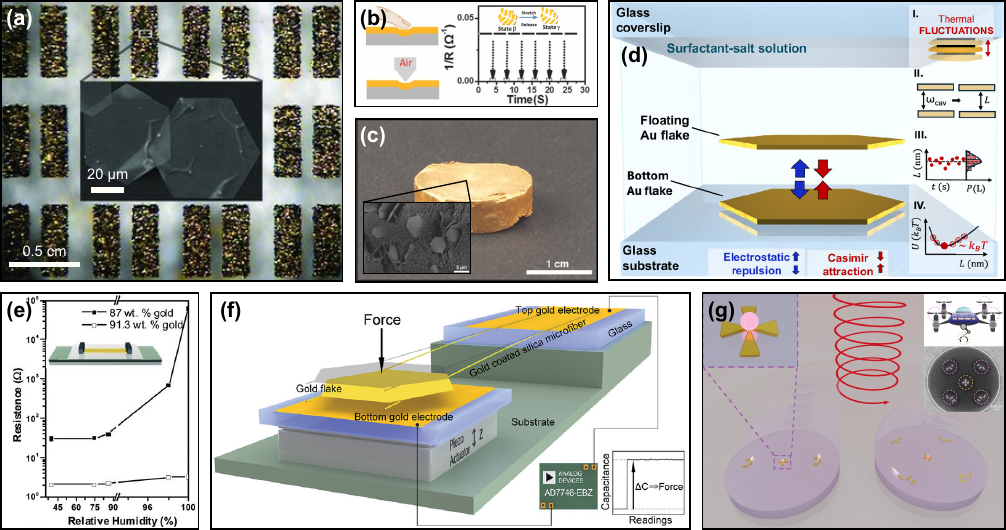}
    \caption{Selected application examples of monocrystalline flakes in nano- and micro-devices. \pnl{a} Micrograph of stretchable and bendable electrodes deposited on \emph{"Ecoflex"} substrates; inset shows close-up SEM image of the indicated area. \pnl{b} \ce{Au} flakes as a force sensor: (right) plot shows variation of in-plane electric resistance of the film upon finger touch or air nozzle blow, insets illustrate applied stimuli, and (right) schematically depict deformation of the sample upon applied mechanical force. \pnl{c}~Photograph of an amyloid fibril-based ultralow-density aerogel; inset shows close-up SEM image revealing porous composition of the sample. \pnl{d} Casimir self-assembly of a pair of \ce{Au} flakes used for force sensing in microfluidics. Panel insets (right) show the main concepts of how thermal fluctuations provide experimental insight into the system’s interaction potential: I) thermal fluctuations of the upper floating flake, II) measurement of separation distances, III) statistical analysis, IV. potential reconstruction. \pnl{e} Amyloid fibril-coated \ce{Au} flake as humidity sensor. Plot shows dependence of the resistance on relative humidity for the films with different \ce{Au} compositions. Inset shows schematic illustration of the system: flakes transferred on a plastic tape with glass-iron bilayer slides as supporters. \pnl{f} Schematic illustration of a miniature capacitive picobalance utilizing \ce{Au} flakes. \pnl{g} Light-driven plasmonic microrobots for nanoparticle manipulation fabricated out of \ce{Au} flake. Insets (right, top) shows an artistic impression depicting the main idea of the microrobot concept and (right, bottom) SEM image of the fabricated microrobot.
    \pnl{a} Reprinted with permission from Ref.~\cite{Moon:2013}, \pnl{b} from Ref.~\cite{Zhou:2015}, \pnl{c}~from Ref.~\cite{Nystrom:2016}, \pnl{d} from Ref.~\cite{Hoskova:2025}, \pnl{e} from Ref.~\cite{Li:2013a}, \pnl{f} from Ref.~\cite{Zhu:2025}, \pnl{g}~from Ref.~\cite{Qin:2025}.
    \label{fig:Applications_ElecMech}}
\end{figure}

Across microelectronic industry, gold is a material of choice for fabrication of electrical contacts and electrodes due to its excellent conductivity and environmental stability. Although pure bulk \ce{Au} is a very ductile malleable material, PVD \ce{Au} thin films can still be fragile~\cite{Augis:1979}, which is not suitable for devices subjected to stretching or bending. Moon~\emph{et~al.} demonstrated a promising alternative: films of conglomerated \ce{Au} flakes [shown in Fig.~\ref{fig:Applications_ElecMech}\pnl{a}] function as highly stretchable and bendable electrodes with excellent electrical conductivity~\cite{Moon:2013}. 
Interestingly, a similar arrangement of disordered \ce{Au} flakes can be used as a pressure (or strain) sensor, as shown in Fig.~\ref{fig:Applications_ElecMech}\pnl{b}: Zhou \emph{et~al.} demonstrated a variation of in-plane electric resistance of the film upon external mechanical force stimulus, like finger touching or air-nozzle blow~\cite{Zhou:2015}. Alternatively, \ce{Au} flakes incorporated into an amyloid fibril-based ultralow-density aerogel [Fig.~\ref{fig:Applications_ElecMech}\pnl{c}] also can be used as resistive pressure sensor~\cite{Nystrom:2016}. In fact, the sensitivity of \ce{Au} flake-based force sensors can be reduced down to the piconewton (pN) level, as demonstrated by Zhu \emph{et~al.}, who developed picobalances with a detection limit of \SI{6.9}{\pico\newton}~\cite{Zhu:2025}. The schematic illustration of such a picobalance device is shown in Fig.~\ref{fig:Applications_ElecMech}\pnl{f}: the response can be precisely calibrated by applying pN-level radiation pressure to a highly reflective \ce{Au} flake surface, and various types of pollen -- spanning masses in the nanogram range (4.6 to \SI{96.3}{\nano\gram}) -- can be readily measured using a picobalance at the single-particle level.
Furthermore, Fig.~\ref{fig:Applications_ElecMech}\pnl{e} shows an example of \ce{Au} flake-nanofibril composite-based humidity sensor: several works~\cite{Li:2013a,Li:2016c,Chen:2017} demonstrated variation of the in-plane resistance of amyloid fibril-coated \ce{Au} flakes due to changes of the ambient relative humidity.
Besides, \ce{Au} flakes can be used for force measurement in microfluidic systems: Ho\v{s}kov\'{a} \emph{et~al.} studied nanoscale surface interactions (Casimir--Lifshitz electrostatic potential) between a pair of \ce{Au} flakes self-assembly in liquid [see Fig.~\ref{fig:Applications_ElecMech}\pnl{d}], highlighting promises for the precise measurement of the ionic strength of liquid and surface charges~\cite{Hoskova:2025}.

\FloatBarrier

Gold flakes were also shown to be useful in manipulation of microparticles: photo- and thermophoretic forces induced upon illumination of the \ce{Au} flakes with a laser beam allows to assemble colloidal microparticles in the vicinity of sharp corners~\cite{Sharma:2020}. In the context of optical tweezers, Yan \emph{et~al.} demonstrated interferometric optical traps created from a single laser beam and its reflection from individual \ce{Au} nanoplates~\cite{Yan:2014}. Kullock \emph{et~al.} have demonstrated the dielectrophoretic manipulation of metallic nanoparticles using an antenna configuration~\cite{Kullock:2020}. More recently, impressive demonstrations of nanoscale miniaturization of modern widespread gadgets, such as light-driven micro-drones~\cite{Wu:2022}, microbots~\cite{Qin:2025} and even robotic cleaners~\cite{Qin:2026}, were demonstrated for nanoparticle manipulation.
Fig.~\ref{fig:Applications_ElecMech}\pnl{g} shows FIB-fabricated nanostructures with either two or four "motors" and a pair of "tweezers", are actuated in aqueous environment using an unfocused circularly polarized laser beam. Such microbots can be made to follow a specific path and capture a single suspended nanoparticle. 
Last but not least, Radha \emph{et~al.} demonstrated that, using electromigration, \ce{Au} flakes can be employed to electrically contact standing indium arsenide (\ce{InAs}) nanowires of uneven heights~\cite{Radha:2012a}; after placing the flake on top of the wires, the shorter ones can be additionally connected by forming electromigrated tips.
This is of particular importance for nanoscale superconducting and quantum devices, as \ce{InAs} semiconductor nanowires exhibit high electronic mobility and a narrow bandgap, making them well suited for the fabrication of Josephson junctions.

\subsection{Plasmonic nanoantennas, waveguides and metasurfaces}
\label{sec:PlasmonicNanoantennas}
\begin{figure}[ht!]
    \centering
    \includegraphics[width=.9\linewidth]{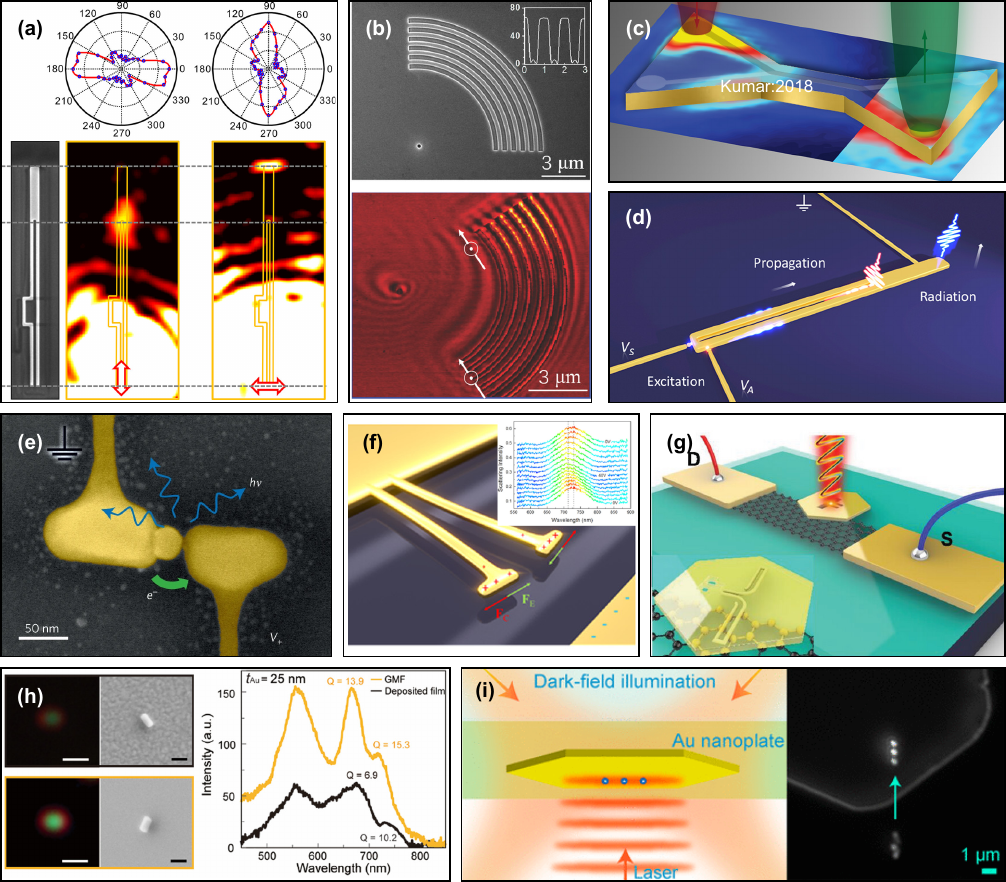}
    \caption{Selected application examples of monocrystalline flakes in nano- and micro-devices. \pnl{a} Plasmonic two-wire transmission lines for mode conversion: SEM image (left), scattering diagrams (top) and far-field scattering images (bottom) for orthogonal polarization excitations. \pnl{b} SEM image (top) and SNOM scan (bottom) of broadband plasmonic logic gates based on curved gratings. \pnl{c} Schematic illustration of an on-chip all-optical information processing device: directional signal transmittance mediated by 2D plasmonic eigenmodes supported by crystalline cavities. \pnl{d} Electrically-modulated two-wire transmission line allows excitation of distinct propagating modes. \pnl{e} Pseudo-colored SEM image of electrically driven optical antennas: lateral tunnel structure for tunable electroluminescence. \pnl{f} Electromechanically tunable suspended optical nanoantenna functioning as a nanoelectrometer; inset shows variations of the scattering spectrum as a function of applied voltage \pnl{g}~Polarization-dependent SPP near-field photodetector based on graphene: SPP waves excited by the excitation laser are absorbed in the graphene layer and converted into electrical signals. \pnl{h}~Comparison of nanoparticle-on-mirror configurations implemented on polycrystalline (top, black line) and monocrystalline (bottom, yellow line) substrates. Graph on the right shows scattering spectra for the two cases. \pnl{i} Enhanced optical trapping and optical binding of \ce{Ag} nanoparticles in interferometric optical traps created from a single laser beam and its reflection from \ce{Au} flake. Illustration on the left hand side shows experimental configuration and image on the right hand side shows experimental dark-field micrograph. \pnl{a} Reprinted with permission from Ref.~\cite{Dai:2014a}, \pnl{b} from Ref.~\cite{Sang:2018}, \pnl{c} from Ref.~\cite{Kumar:2018}, \pnl{d} from Ref.~\cite{Ochs:2021}, \pnl{e} from Ref.~\cite{Kern:2015}, \pnl{f} from Ref.~\cite{Chen:2016}, \pnl{g} from Ref.~\cite{Wu:2023a}, \pnl{h}~from Ref.~\cite{Liu:2022}, and \pnl{i} from Ref.~\cite{Yan:2014}.
    \label{fig:Applications_Plasmonics}}
\end{figure}

The use of crystalline metal is beneficial for nanooptics and plasmon-based applications for several reasons. First, the well-defined crystallinity allows precise top-down fabrication of high-definition nanostructures with fine features over a large extend, such as plasmonic two-wire transmission lines with subwavelength gaps extending several micrometers~\cite{Dai:2014a}, shown in Fig.~\ref{fig:Applications_Plasmonics}\pnl{a,d}. Such high-quality waveguides enable the deterministic selective excitation and modulation of guided plasmonic modes by electricity~\cite{Ochs:2021}. 
Other demonstrations of crystalline plasmonic waveguides and integrated circuit components include multimode plasmonic two-wire transmission lines~\cite{Geisler:2013}, directional coupler nanocircuits~\cite{Razinskas:2016,Rewitz:2014}, spin sorting devices~\cite{Krauss:2019}.
Geisler \emph{et~al.} reported Fabry--P{\'e}rot oscillations in the transmission of plasmons through a nanowire~\cite{Geisler:2017}, paving the way towards ultra-compact interferometric devices.

Secondly, as discussed in Sec.~\ref{sec:OpticalProperties}, the absence of surface corrugation and crystal boundaries somewhat reduces the scattering loss of propagating SPP, especially in configurations with strong field confinement and large propagation wave vectors, which increase susceptibility to roughness. Figure~\ref{fig:Applications_Plasmonics}\pnl{b} shows an example of a device that exploits this feature: curved gratings with a single hole output fabricated on monocrystalline \ce{Ag} flakes function as broadband Boolean logic devices that achieve a high intensity contrast ratio up to 29\,dB~\cite{Sang:2018}. Since the signal relies on the free propagating SPPs, the ultrasmooth surface and grain-free crystalline nature of the monocrystalline \ce{Ag} flakes plays a critical for the high contrast ratio. 
This aspect is also of crucial importance for imaging localized and propagating plasmon modes supported by nanoantennas using, e.g., photoelectron emission microscopy (PEEM). For example, this technique allowed to study photoemission polarization~\cite{Klaer:2016} and angular momentum~\cite{Klaer:2015} in plasmonic cross antennas, and near-fields imaging with high spatial resolution~\cite{Razinskas:2016} of propagating modes in plasmonic nanocircuits. In these cases high signal-to-noise ratio was also ensured by the reduced scattering from crystalline surfaces.
The superior quality of monocrystalline \ce{Au} flakes further facilitates the realization of nanostructures supporting 2D plasmonic eigenstates for directional signal routing and the creation of Boolean logic gates, such as shown in Fig.~\ref{fig:Applications_Plasmonics}\pnl{c}~\cite{Viarbitskaya:2013, Kumar:2018, Kumar:2021, DellOva:2024}.

Apart from plasmonic waveguides, crystalline metal flakes were extensively used for the fabrication of various nanoantennas (some examples were already mentioned in Sec.~\ref{sec:NanoMicroFab}). 
An important development in this direction was demonstrated by Prangsma \emph{et~al.}, who proposed electrically connected resonant optical antennas fabricated out of crystalline \ce{Au}~\cite{Prangsma:2012}. Such a configuration, shown in Fig.~\ref{fig:Applications_Plasmonics}\pnl{e}, was proven to be versatile for many applications. 
For example, it can be used as a nanoscale light source, as showcased by Kern \emph{et~al.}: electron tunneling across the gap gives rise to electroluminescence~\cite{Kern:2015}, which is tunable by the antenna geometry and applied voltage; however, such tunneling gaps cannot yet be fabricated by FIB and are instead typically realized via electromigration.
Building on that demonstration, Grimm \emph{et~al.} have shown that filling the antenna gap with the organic semiconductor zinc phthalocyanine (\ce{ZnPc}) enables tunable (emission wavelength- and directionality-switchable) organic light-emitting devices~\cite{Grimm:2022}. Other examples of crystalline electrically driven optical antennas include Yagi--Uda antennas~\cite{Kullock:2020}. 
Notably, essentially the same geometry, but with antennas embedded into a wide-bandgap semiconductor like titanium dioxide (\ce{TiO2}) works as a tunable nanoplasmonic photodetector, as demonstrated by Pertsch \emph{et~al.}~\cite{Pertsch:2022}. Owing to suppression of electron-phonon scattering and improved hot carrier tunneling efficiency, monocrystalline  nanowire-based devices represent one of the highest reported photovoltage sensing and hot carrier collection performances in terms of on-chip device density and responsivity per area~\cite{Zhu:2025a}. Alternative configurations of nanoscale plasmonic photodetectors were also demonstrated by interfacing \ce{Au} flakes with graphene [see Fig.~\ref{fig:Applications_Plasmonics}\pnl{g}]~\cite{Wu:2023a}. 
Additionally, similar geometry can be used for electrically-driven site-selective functionalization (optimization of interfacial charge-carrier injection)~\cite{Ochs:2023}.

Indeed, employing crystalline metal as a source material for FIB or EBL fabrication of plasmonic nanostructures yields high-fidelity results~\cite{Mejard:2017,Kejik:2020}, yet it still relies on the precision of the machinery and posses drawbacks like contamination with \ce{Ga} ion. In turn, using pristine metal flakes as a perfect mirror for the creation of a nanoparticle-on-mirror configuration [see Fig.~\ref{fig:Applications_Plasmonics}\pnl{h} and \pnl{i}] avoids those issues. These cavities support so-called gap plasmon modes, with extremely small (single digit \si{\nano\meter\cubed} or even sub- \si{\nano\meter\cubed} scale) mode volumes, which allow to enhance linear and nonlinear light-matter interaction by orders of magnitude~\cite{Chikkaraddy:2016,Jakob:2023} (this feature is particularly useful for biological and chemical sensing and quantum emitter enhancement, which will be discussed in the following subsections).
Here, ultrasmooth surfaces of crystalline flakes are extremely useful to create robust and reproducible nanoparticle-on-mirror cavities. For example, Wang \emph{et~al.} demonstrated significant effect of the substrate quality on the optical response of such nanocavities~\cite{Wang:2023a}, paving the way towards low-loss plasmonics~\cite{Liu:2022} and improved optical trapping~\cite{Yan:2014}.

\FloatBarrier

In the context of metasurfaces, Neal \emph{et~al.} reported a seeded growth of large-area, epitaxially aligned arrays of single-crystal \ce{Au} nanotriangles with atomically flat surfaces and well-defined plasmonic modes, advancing their integration into wafer-based nano-enabled devices~\cite{Neal:2022}.
Boroviks \emph{et~al.} demonstrated a gap plasmon based metasurface for beam steering, which utilized a thick crystalline \ce{Au} flake as a substrate~\cite{Boroviks:2019}. The performance of this device showed a modest, yet measurable (approximately 5\%), efficiency improvement for visible light in comparison with a polycrystalline counterpart. Also, epitaxially-grown \ce{Au} monocrystalline substrates allow to fabricate metasurfaces with high quality and reproducibility~\cite{Grayli:2024}

\subsection{Biological and chemical sensing}

Gold and its derived nanomaterials are known to be chemically inert and biocompatible~\cite{Shukla:2005,Lewinski:2008}, and hence have found numerous applications in chemistry, biology and biomedicine~\cite{Dreaden:2012}. In this regard, crystalline flakes are no exception in the plethora of \ce{Au} colloids, and have also been exploited for enhancing sensing, catalytic reactions and other biochemical processes, which we review in this section.

\begin{figure}[ht!]
    \centering
    \includegraphics[width=.9\linewidth]{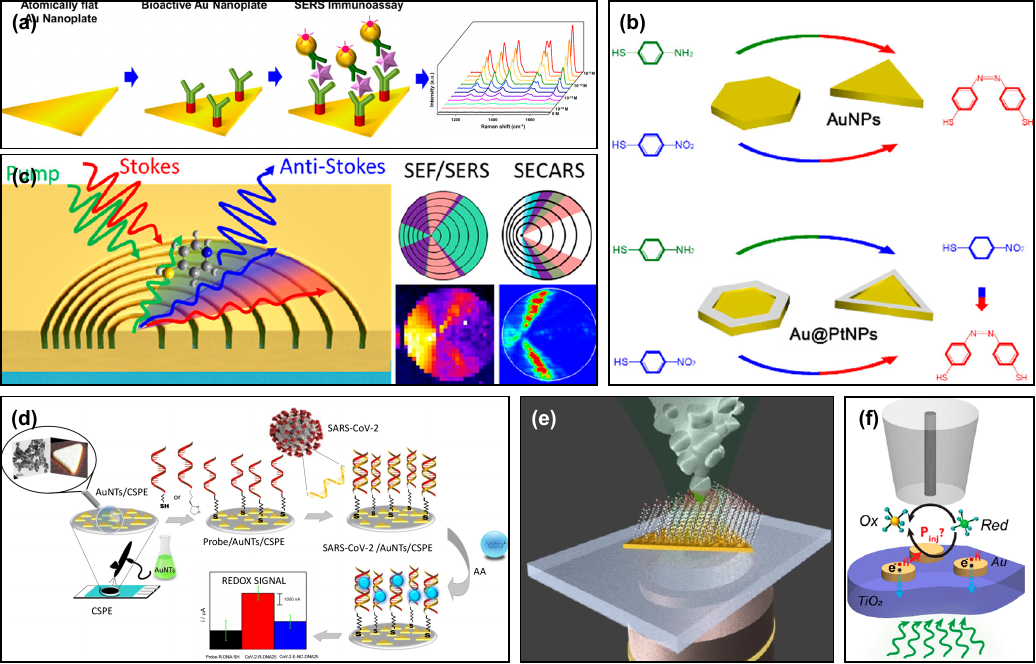}
    \caption{Selected application examples of monocrystalline flakes in bio and chemical sensing, and catalysis.
    \pnl{a} Ultraspecific attomolar detection of C-reactive protein biomarkers. Flowchart shows immobilization process and graph on the right hand-side show SERS spectra of rhodamine B isothiocyanate, which is bound to the \ce{Au} nanoparticles, and serves as a marker (SERS nanotag). \pnl{b} Photocatalytic conversion of 4-aminothiophenol and 4-nitrobenzenethiol on bare \ce{Au} flakes (top) and on \ce{Pt}-coated flakes (bottom). \pnl{c}~Plasmonic Doppler gratings for surface-enhanced coherent anti-Stokes Raman scattering. \pnl{d} Detection of SARS-CoV-2 DNA and RNA sequences using \ce{Au} flakes functionalized with oligonucleotides. \pnl{e} TERS experimental setup for the measurement of near-field temperature experienced by self-assembled 16-mercaptohexadodecanoic acid monolayers covalently bond to the \ce{Au} flake surface. \pnl{f}~Schematic illustration of a plasmonic photocatalytic device: metal/semiconductor heterostructure for interfacial hot carrier generation, transport and collection. \pnl{a}~Reprinted with permission from Ref.~\cite{Hwang:2019}, \pnl{b}~from Ref.~\cite{Li:2020a}, \pnl{c}~from Ref.~\cite{Ouyang:2021}, \pnl{d} from Ref.~\cite{DelCano:2022}, \pnl{e}~from Ref.~\cite{Richard:2020}, and \pnl{f} from Ref.~\cite{Kiani:2023}.
    \label{fig:Applications_Chemical}}
\end{figure}

The prime example of such applications is surface-enhanced Raman spectroscopy (SERS)~\cite{Langer:2020}. Since conventional Raman (inelastic light) scattering is an optical process with extremely low efficiency, its enhancement in the vicinity of metallic surface is considered as one (if not the only) of the practical applications of plasmonics~\cite{Khurgin:2015c}, especially in the nanoparticle-on-mirror configuration~\cite{Zhu:2014, Peng:2024}. 
Conventionally, rough metallic substrates are preferred for SERS applications since, as surface roughness gives rise to uniformly distributed plasmonic hotspots -- regions of intense electric field enhancement and high local density of photonic states that enhance the Raman scattering. 
At first glance, atomically smooth crystalline flake surfaces may seem unsuitable for these purposes. However, when spatially localized Raman signals are desired, i.e., SERS from a specific location, monocrystalline flakes have proven to be useful: sharp edges and corners of pristine flakes, nanopatterned features, or nanoparticles placed atop them allow the creation of well-defined plasmonic hotspots.
In turn, flat, unpatterned facets do not produce any plasmonic field enhancement, thereby introducing very low background noise. Furthermore, well-defined crystallographic facets allow selective chemical binding, thereby increasing specificity of the sensing.
Hence, over the recent years, \ce{Au} flakes have proven to be a reliable SERS substrates~\cite{Sun:2007,Biagorri:2008, Xia:2013, Nootchanat:2013, Lin:2014a, Zhou:2015a, Chen:2016a, Fang:2016, Gwo:2016, Xin:2017, Lv:2018, Chen:2019b, Wang:2020}. Among these examples, the demonstration of ultraspecific attomolar (aM) detection of protein biomarkers by Hwang \emph{et~al.} stands out~\cite{Hwang:2019}. In this work, specificity and highly improved detection limit of 10\,aM were achieved by multi-step preparation of the \ce{Au} flake substrate, immobilizing them with anti-C-reactive protein and attaching rhodamine B isothiocyanate-coated nanoparticle atop to create a nanoparticle-on-mirror geometry, as illustrated in Fig.~\ref{fig:Applications_Chemical}\pnl{a}. 
Also, nanopatterned \ce{Au} flakes have shown substantial enhancement of Raman signal, e.g., Sweedan \emph{et~al.} have reported a novel SERS platform based on evolutionary-optimized double wire gratings with extremely fine features~\cite{Sweedan:2024}. 

Another widespread application of \ce{Au} flakes is related to electrochemistry, where the lack of crystal boundaries, in some cases can also lead to more effective reactions due to facilitated transport of high-energy electrons in the crystal. An example of enhanced catalysis of 4-aminothiophenol and 4-nitrobenzenethiol molecules using bare and \ce{Pt}-coated \ce{Au} flakes is shown in Fig.~\ref{fig:Applications_Chemical}\pnl{b}~\cite{Li:2020a}. Furthermore, the $\{111\}$ crystal facet offers ballistic collection of high-energy $d$-band holes, which is only possible in nanoantennas fabricated from monocrystalline flakes that do not contain grain boundaries that would otherwise cause carriers to lose energy through scattering. Such unique facet-dependent electrochemistry is only available in nanostructures fabricated from monocrystalline \ce{Au} substrates~\cite{Kiani:2023}, as demonstrated by Kiani \emph{et~al.} [see Fig.~\ref{fig:Applications_Chemical}\pnl{f}]. 
Other examples of catalytic action enhanced by the presence of crystalline metal flakes includes methanol oxidation~\cite{Li:2009,Yang:2019c}; oxidation of dopamine and ascorbic acid~\cite{Goyal:2009}; formic acid oxidation~\cite{Momeni:2016}; \ce{H2O2} decomposition~\cite{He:2018} and subsequent glucose sensing~\cite{Zhang:2011}; catalysis reduction of 4-nitrophenol to 4-aminophenol~\cite{Ma:2022}, reduction of methyl red and blue dyes~\cite{Bhosale:2016}, and oxidation of \ce{Fe(CN)6^4-}~\cite{Sabzehparvar:2026}.

Over the recent years, \ce{Au} flakes have been used to create a spectrometer-free imaging-based sensing platform, namely plasmonic Doppler gratings (PDG)~\cite{See:2017}, shown in Fig.~\ref{fig:Applications_Chemical}\pnl{c}, which provide continuous azimuthal angle-dependent periodicity for image-based index sensing applications. Due to the atomically flat surface of the \ce{Au} flakes, this platform enables spectrometer-free index sensing with extremely low background noise~\cite{Lin:2019}. The monocrystalline PDG platform has also been used in plasmon-enhanced nonlinear spectroscopy, where multiple input and output frequencies are involved. By performing hyperspectral mapping, the best grating periodicity that simultaneously enhances the input and output frequencies can be identified. Since various combinations of enhancement effects can be obtained in one image, PDG provides a simple yet effective platform for the investigation of the enhancement effect. For example, PDG was used to study the enhancement mechanism in surface-enhanced coherent anti-Stokes Raman scattering (SECARS), which is a third-order nonlinear optical process involving three input frequencies (pump, probe, and Stokes) and one output frequency (anti-Stokes signal)~\cite{Ouyang:2021}. The azimuthal distribution of the signal enhancement provides rich information on the enhancement mechanism. The identification of the spatial distribution relies on a good signal-to-noise ratio, therefore, the use of monocrystalline PDG is critical. 
Similar image based analysis was performed for other nonlinear processes like second-harmonic generation (SHG) and two-photon photoluminescence (TPPL) to understand the difference between these multi-photon processes and to see the effect of surface lattice resonance in four-wave mixing (FWM)~\cite{Barman:2022, Chakraborty:2023}.

Remarkably, \ce{Au} flakes have already found applications in bio-medicine, as a novel drug carrier platform for cancer therapy~\cite{Singh:2014, Brann:2016} and detection~\cite{He:2018, Eom:2021}. Recently, they have been also exploited for the detection of notorious SARS-CoV-2 coronavirus or its antigens~\cite{DelCano:2022, Yue:2022, Wu:2022a, Xu:2023}, using SERS or electrochemical measurement of reduction–oxidation reaction, as shown in Fig.~\ref{fig:Applications_Chemical}\pnl{d}.

In-vivo biological and diagnostic applications have similarly benefited from these crystals; for instance, movable \ce{Au} microplates serve as non-toxic, "cytophilic" single-cell platforms where atomic smoothness enhances fluorescence signals by an order of magnitude~\cite{Radha:2010}.

Owing to their ultrasmooth surfaces, crystalline flakes are excellent substrates for studying monolayer absorbents~\cite{Egelbrecht:1983}. For example, Boya \emph{et~al.} showed that a capacitor formed by \ce{Au} flake and substrate metal is capable of electrical analysis of molecular monolayer adsorbants (in particular alkanemono- and -dithiols)~\cite{Boya:2013}, whereas Jeong \emph{et~al.} demonstrated similar electrical analysis of hexadecanethiol monolayers adsorbed to a a flake using conductive probe AFM~\cite{Jeong:2016}. Monocrystalline \ce{Ag} plates were used as a substrate for detection and electrochemical analysis of mono- and multilayer adsorbents using tip-enhanced Raman scattering (TERS)~\cite{Deckert-Gaudig:2009}, as illustrated in Fig.~\ref{fig:Applications_Chemical}\pnl{e}. More on TERS experiments with \ce{Au} flakes and other scanning probe microscopies experiments are discussed in Sec.~\ref{sec:IdealMirrors}. 

\FloatBarrier

\subsection{Applications in quantum plasmonics and enhancement of single-photon sources}

Monocrystalline flakes also serve as promising platforms for quantum nanophotonics, in particular for enhancement of the emission rate of single photon sources, and their coupling to plasmonic waveguides, which are envisioned as crucial components for future quantum technologies~\cite{Bozhevolnyi:2017a}. 

It is well known that the spontaneous emission rate of molecules and other quantum emitters is drastically enhanced in the vicinity of a reflecting interface, although in a very close proximity to the surface the radiative decay is quenched~\cite{Drexhage:1966,Drexhage:1968}. Hence, crystalline flakes, due to their atomic-level flatness, act as nearly perfect mirrors, allowing to mitigate non-radiative losses due to surface roughness. For example, Song \emph{et~al.} reported approximately 16-fold photoluminescence enhancement from a single quantum dot placed atop of an \ce{Au} flake (relative to the one placed atop of a glass substrate)~\cite{Song:2014}. 
It is important to note that the distance between the metal surface and the quantum dot should be carefully controlled, as at very small separations the emitter can experience \emph{"quenching"} -- reduced apparent quantum yield due to coupling to nonradiative modes~\cite{Goncalves:2020}.

Not being limited by a flat surface geometry, the emission rate can be enhanced even further by making use of plasmonic nanostructures, which, within the quantum nanophotonics community, are often referred to as plasmonic cavities~\cite{Russel:2012}. In the weak cavity-emitter coupling regime, which is relatively common for plasmonic systems~\cite{Tserkezis:2023}, the main physical mechanism behind this enhancement is the so-called \emph{"Purcell effect"}~\cite{Pelton:2015,Goncalves:2020}. The Purcell factor $F_\mathrm{P}$ quantifies the enhancement of the photonic density-of-states (relative to vacuum) and in a cavity context with spectrally well-separated modes, it is directly proportional to the mode quality factor $Q$ and inversely proportional to the mode volume $V$, i.e., $F_\mathrm{P}\propto Q/V$. In fact, generation of quantum light is considered as one of the few truly useful applications of plasmonics: despite intrinsic losses in metal, the Purcell effect can be so strong that these non radiative losses are overcome~\cite{Bozhevolnyi:2017b, Fernandez-Dominguez:2018}.
Thus, advantages offered by monocrystalline plasmonic antennas and waveguides, which were discussed in Sec.~\ref{sec:PlasmonicNanoantennas}, 
are also beneficial for quantum plasmonics applications: both extremely small mode volumes and increased $Q$-factors (in comparison with polycrystalline counterparts) contribute to increased $F_\mathrm{P}$ values.

\begin{figure}[ht!]
    \centering
    \includegraphics[width=.9\linewidth]{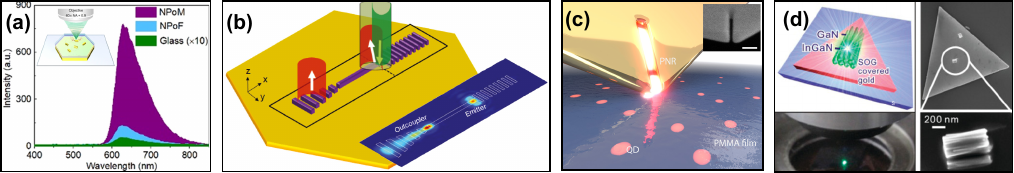}
    \caption{Selected application examples of monocrystalline flakes in quantum nanophotonics. \pnl{a}~Enhancement of fluorescence from crystal violet molecules embedded in a gap between \ce{Au} nanorod and crystalline flake. The plot shows emission spectra on glass, polycrystalline and crystalline substrates; inset shows illustration of the experiment. \pnl{b}~A quantum emitter (\ce{Ge} vacancy in a nanodiamond) coupled to a hybrid plasmonic waveguide. \pnl{c}~Schematic illustration and SEM image (top right corner) of a broadband plasmonic nanoresonator probe for interacting with quantum dots embedded in a polymer film. \pnl{d}~Plasmonic nanolaser based on \ce{InGaN}/\ce{GaN} semiconductor nanorods. Schematic (top left) shows arrangement of the device, photograph (bottom left) shows green light emission from the device sample and SEM images (right) show the fabricated device. \pnl{a} Reprinted with permission from Ref.~\cite{Zhang:2024}, \pnl{b} from Ref.~\cite{Siampour:2020}, \pnl{c} from Ref.~\cite{Gross:2018}, and \pnl{d}~from Ref.~\cite{Wu:2011}.
    \label{fig:Applications_Quantum}}
\end{figure}

In this regard, nanoparticle-on-mirror cavities are particularly useful, as they enable quantum emitters to be embedded within the plasmonic gap~\cite{Baumberg:2019}. Nanometer~\cite{Chikkaraddy:2016} (or even sub-nanometer~\cite{Benz:2016})-scale mode volumes allows to achieve photon sources that are particularly bright and with fast decay rates~\cite{Hoang:2015}. However, nanocavities realized on a polycrystalline substrates suffer from poor reproducibility and sample-to-sample irregularities, as in this case surface roughness may manifest on length scales comparable to the gap size. Here, atomic flatness offered by the crystalline metal flakes is especially valuable~\cite{Peng:2024}. 
In the example shown in Fig.~\ref{fig:Applications_Quantum}\pnl{a}, Zhang \emph{et~al.} demonstrated a 143-fold fluorescence enhancement and a 5-fold lifetime reduction for crystal violet molecules embedded in a PMMA layer placed between an \ce{Au} flake and nanorod~\cite{Zhang:2024}. Kumar \emph{et~al.} showed that the decay rate enhancement factor reaches value of $\sim 50$ for the nitrogen-vacancy centers placed in a plasmonic gap between an \ce{Ag} nanowire and an \ce{Ag} flake~\cite{Kumar:2019}. Similarly, rare-earth-metal-doped upconversion nanoparticles placed in a gap between an \ce{Ag} nanocube and an \ce{Ag} flake leads to a drastic acceleration of emission, simultaneously allowing control over polarization and directionality~\cite{Chen:2022}.
Use of dielectric particles atop of crystalline substrates was also proven to be beneficial: Lin \emph{et~al.} demonstrated the influence on the emission spectrum of fluorescent semiconductor microspheres by placing them on an \ce{Au} flake~\cite{Lin:2023a}. They attributed the observed spectral change to alternation of the whispering gallery modes supported by microspheres due to presence of a substrate. Here, monocrystalline metallic substrates are essential for identifying the effect of the metal on transverse-electric and transverse-magnetic whispering-gallery modes: a polycrystalline substrate with significant surface roughness would not allow one to clearly isolate the enhancement due to the metallic mirror, since these modes can couple to SPP modes through surface-induced scattering.

Apart from the emission enhancement due to the Purcell effect, quantum emitters may be also coupled to high-quality crystalline plasmonic waveguides. This allows to delocalize and redirect quantum emission: as shown in Fig.~\ref{fig:Applications_Quantum}\pnl{b}, a hybrid (dielectric-loaded) plasmonic waveguide fabricated atop of an \ce{Au} flake, with embedded germanium-vacancy nanodiamond, allows to separate excitation and emission spots by several micrometers, while preserving single-photon source properties~\cite{Siampour:2020}. Similarly, dibenzoterrylene molecules in anthracene nanocrystals embedded into plasmonic V-grooves milled in \ce{Au} crystals, couple with propagating plasmon modes, and upon out-coupling to the far-field retain quantum light source properties at distances up to $\sim \SI{15}{\micro\meter}$~\cite{Kumar:2020}. 
In another example, Kumar \emph{et~al.} demonstrated the propagation and sorting of single plasmons launched on a crystalline \ce{Au} flake surface by broadband quantum emission from a nanodiamond hosting a single nitrogen–vacancy center~\cite{Kumar:2020a}.
Also, several works have reported coupling single quantum emitters to plasmonic wire waveguides fabricated from crystalline metal flakes~\cite{Kumar:2013, Wan:2015, Schorner:2019, Schorner:2020}.

So far, we have primarily focused on the Wigner–Weisskopf regime, in which the emitter is weakly coupled to the local photonic density of states of its environment and the decay of the emitter is exponential in time. As the Purcell enhancement increases such that the coupling strength exceeds the losses, the system may enter the strong-coupling regime with Rabi-like oscillations in time~\cite{Torma:2015,Tserkezis:2020a,Tserkezis:2023}. As an example exploiting crystalline \ce{Au} flakes, Gro{\ss} \emph{et~al.} demonstrated deterministic strong coupling at room temperature between a quantum dot and a plasmonic gap resonator positioned at the apex of a SNOM tip~\cite{Gross:2018}, later also inferred in a photon-correlation experiments~\cite{Friedrich:2025}. As illustrated in Fig.~\ref{fig:Applications_Quantum}\pnl{c}, the tip features a slit fabricated by FIB milling in a crystalline \ce{Au} flake.

\FloatBarrier

Finally, interfacing crystalline \ce{Au} flakes with a gain medium opens prospects for plasmonic nanolaser devices, which were demonstrated by Wu \emph{et~al.} using indium gallium nitride and gallium nitride (\ce{InGaN}/\ce{GaN}) semiconductor nanorods~\cite{Wu:2011}. The schematics and experimental images of the device are shown in Fig.~\ref{fig:Applications_Quantum}\pnl{d}.
The importance of the substrate quality for such plasmonic lasers was further highlighted by Lee~\emph{et~al.} who showed that using \ce{Au} flakes or epitaxially grown \ce{Ag} films can significantly reduce the lasing threshold (down to \SI{0.2}{\mega\watt\per\centi\meter\squared}) in indium gallium arsenide phosphide (\ce{InGaAsP}) quantum well-based device prototypes~\cite{Lee:2017}. Furthermore, as demonstrated by Chou~\emph{et~al.}, using monocrystalline  aluminum (\ce{Al}) in such plasmonic semiconductor lasers permits operation at high temperatures (up to \SI{353}{\kelvin})~\cite{Chou:2018}. Such device improvements -- namely the increased operating temperature and reduced lasing threshold -- are promoted by reduced losses in the crystalline substrates, since lasing occurs only when the cavity losses are compensated~\cite{Oulton:2009,Repp:2025,Vitale:2024}. 

\section{Crystalline flakes for nonlinear plasmonics}
\label{Sec:nonlinear}

Since the first experimental observation of second-harmonic generation (SHG) from a metal boundary by Brown and colleagues in 1965~\cite{Brown:1965}, the nonlinear optical response from thin metal films and metallic nanostructures have been extensively studied. Although the intrinsic nonlinearity of bulk metal is relatively weak, it can be significantly enhanced by virtue of the plasmon resonances. This enhanced nonlinear response caused by resonant interaction between incident light and the free electrons in plasmonic nanostructures gave rise to a new field of research -- \emph{nonlinear plasmonics}~\cite{Kauranen:2012}.
Although the plethora of nonlinear phenomena is vast, we focus on the two most widely investigated nonlinear processes at metal surfaces: second-harmonic generation (SHG) and nonlinear photoluminescence (NPL). In particular, we will elaborate on how the nonlinear optical response of the crystalline metals is fundamentally different from that of the polycrystalline counterparts.
We note that other types of nonlinear interactions at metal surfaces -- such as third-harmonic generation~\cite{Lippitz:2005,Boyd:2014}, FWM~\cite{Renger:2011}, and high-harmonic generation~\cite{Kim:2008} -- are also present in the crystalline metal flakes, but have so far remained largely unexplored, with only a few reports briefly addressing, for example, THG~\cite{Pan:2024} and nonlinear absorption~\cite{Chen:2013}.

\subsection{Second-harmonic generation}

Second-harmonic generation, also known as frequency doubling, is a type of second-order nonlinear parametric process, in which two photons with energy $\hbar\omega$ combine into a photon with energy $2\hbar\omega$. The strength of this interaction is described by the material-dependent second-order susceptibility tensor $\bm{\chi}^{(2)}$, which relates the excitation electric field $\mathbf{E}(\omega)$ to the nonlinear polarization as $\mathbf{P}^{(2)}(2\omega)=\varepsilon_0 \bm{\chi}^{(2)}:\mathbf{E}(\omega)\mathbf{E}(\omega)$, where $\varepsilon_0$ is the permittivity of the free space, and "$:$" denotes the double dot product~\cite{Shen:2003}. SHG and other even-order nonlinear optical processes require (within the dipole approximation) a non-centrosymmetric medium~\cite{Boyd:2020}. 
Bulk noble metals typically exhibit an FCC crystal structure [as illustrated in Fig.~\ref{fig:CrystalStructure}\pnl{d}], which is centrosymmetric. Nevertheless, the inversion symmetry is naturally broken at any surface, which allows for the \emph{surface} SHG. 
Hence, SHG in noble metals is primarily a phenomenon localized to a surface as $\mathbf{P}^{(2)}_\mathrm{S}(\mathbf{r},2\omega)=\varepsilon_0 \delta(\mathbf{r}-\mathbf{r}_\mathrm{S}) \bm{\chi}^{(2)}_{\mathrm{S}}:\mathbf{E}(\mathbf{r},\omega)\mathbf{E}(\mathbf{r},\omega)$, where $\mathbf{r}_\mathrm{S}$ is the position vector of the surface and $\delta$ is the Dirac delta function. This explains the overall low efficiency of surface SHG from metal films: although the absolute value of $\bm{\chi}^{(2)}_{\mathrm{S}}$ may be relatively large, the light-matter interaction is limited to a vanishingly thin region at the interface, resulting in overall small nonlinear induced polarization.
Bulk-like quadrupolar (and higher order) contributions may also contribute to the response, however they are assessed to be orders of magnitude weaker~\cite{Wang:2009}. Besides, \ce{Au}, \ce{Ag} and \ce{Cu} metals are highly reflective or absorptive at optical frequencies, implying that the interaction volume remains restricted to a near-interface region of approximately 10\,nm thickness, which corresponds to the penetration depth of the incident wave inside metal.
Nevertheless, nonlinear optical processes at metal surfaces can be substantially amplified by virtue of strong field enhancement caused by surface plasmon resonances~\cite{Kauranen:2012,Butet:2015}. 
Both resonances at the fundamental and second-harmonic frequencies are important: the former enhances the excitation field, while the latter improves emission outcoupling~\cite{Yang:2017}. Some examples of such resonant structures will be discussed later in this section.

\begin{figure}[ht!]
    \centering
    \includegraphics[width=.7\linewidth]{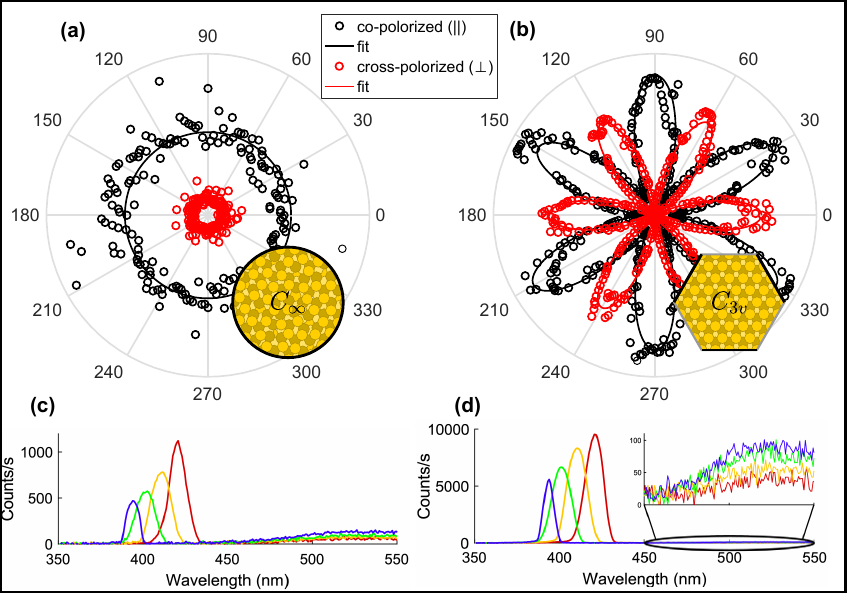}
        \caption{Anisotropic SHG from crystalline \ce{Au} surface. \pnl{a,b} Polar plot of second-harmonic intensity as a function of crystal orientation angle for co-polarized (black) and cross-polarized (red) signals at polycrystalline surface \pnl{a} and $\{111\}$-type surface \pnl{b}. Insets in \pnl{a,b} show schematic diagrams of the first few layers of atom arrangements at the two types of surfaces. \pnl{c,d} Spectrum of the nonlinear reflection from polycrystalline surface \pnl{c} and $\{111\}$-type surface \pnl{d}. The inset in \pnl{d} shows a zoom-in of the measured NPL spectral region (450--550\,nm). Curve colors indicate excitation wavelength: 780\,nm (blue), 800\,nm (green), 820\,nm (yellow), and 840\,nm (red). Reprinted with modifications from Ref.~\cite{Boroviks:2021}.
   \label{fig:SHG}}
\end{figure}

The components of the $\bm{\chi}^{(2)}_{\mathrm{S}}$ tensor for noble metals surfaces is determined by their structural characteristics. In polycrystalline PVD metal films, the crystal domains typically have random orientation and subwavelength characteristic dimensions. Consequently, any anisotropic crystal property averages out, making the films effectively in-plane isotropic.
By symmetry, SHG from such centrosymmetric and isotropic films, can be only a result of interaction with surface-normal $\bm{\chi}^{(2)}_{\mathrm{S}}$ components, such as $\chi^{(2)}_{\perp\perp\perp}$, $\chi^{(2)}_{\perp\parallel\parallel}$, $\chi^{(2)}_{\perp\perp\parallel}$. Hence, the SHG from PVD gold films stems primarily from the surface-normal component of the excitation electric field, which was demonstrated already in the seminal experiments by Brown \emph{et~al.}~\cite{Brown:1965}. Furthermore, as shown in Fig.~\ref{fig:SHG}\pnl{a}, SHG from a polycrystalline gold film is independent of the relative sample orientation: it is primarily co-polarized with the excitation laser polarization, with a minor cross-polarized contribution which is due to experimental artifacts.

The situation changes drastically in the case of crystalline metals, where the surface second-order response is determined by the symmetries of the crystal facets. This results in in-plane anisotropy of SHG from crystalline surfaces, even for the simplest principal types of the FCC crystal facts -- $\{111\}$, $\{110\}$and $\{100\}$. As a prime example, for $\{111\}$-type surface, $\chi^{(2)}_{\parallel\parallel\parallel}$ becomes non-vanishing due to broken centrosymmetry within the first three subsurface atomic layers and is anisotropic relative to the $\langle 11\hat{2}\rangle$ and $\langle\hat{1}10\rangle$ crystal axis [see schematic illustration in Fig.~\ref{fig:CrystalStructure}\pnl{d}]. 
Earlier theoretical analysis of the $\bm{\chi}^{(2)}_{\mathrm{S}}$ tensor properties based on symmetry considerations was done by Sipe \emph{et~al.}~\cite{Sipe:1987}, whereas Boroviks \emph{et~al.} demonstrated experimentally that in the case of $\{111\}$-type surface of \ce{Au} flakes SHG polarization follows a distinct dependence on the relative orientation of the sample and excitation laser polarization~\cite{Boroviks:2021}. The six-fold petal pattern in the intensities of co- and cross-polarized SHG signals is clearly seen in Fig.~\ref{fig:SHG}\pnl{b} (similar observation was later reported for \ce{Cu} flakes by Dayi \emph{et~al.}~\cite{Dayi:2025a}, and the importance of surface-tangential second-harmonic source in crystalline \ce{Au} flakes was also highlighted by Mathew \emph{et~al.}~\cite{Mathew:2025}).
Overall, SHG signal intensity (as measured in a conventional nonlinear microcopy setup) is approximately one order of magnitude stronger for \ce{Au} flake $\{111\}$ surface, as compared with polycrystalline surface [compare Fig.~\ref{fig:SHG}\pnl{c} and \pnl{d}]. In contrast to the thick samples studied in Ref.~\cite{Boroviks:2021}, ultrathin (sub-10\,nm) flakes exhibit a non-monotonic dependence of SHG intensity on thickness~\cite{Pan:2024}. This nontrivial behavior has been attributed to quantum surface effects, namely the quantization of electronic energy levels in samples comprising only a few atomic layers. As a result, optical transitions between intersubband levels become thickness-dependent and display pronounced oscillatory features (also discussed in the context of third-order nonlinear effects~\cite{Qian:2016, Grossmann:2019, RodriguezEcharri:2021b}).

The above discussed features highlight that the nonlinear optical processes are particularly sensitive to crystallinity and can effectively probe the crystallographic orientation of materials at the nanoscale. 
The polarization-dependent response not only underscores the potential of second-harmonic microscopy as a robust, non-destructive technique for characterizing crystallographic orientation but also provides valuable insights for enhancing nonlinear responses in plasmonic systems, ultimately paving the way for advanced applications in nanophotonics and optoelectronics. However, so far the anisotropic nature of SHG was directly exploited only in a few recent works~\cite{Chen:2024c,Boroviks:2025}. For example, Boroviks and Martin have explored the possibility to selective enhance (or silence) the SHG signal stemming from anisotropic and isotropic $\bm{\chi}^{(2)}_{\mathrm{S}}$ tensor components. Using LSP resonance supported by the FIB-milled grooves, they showed control over co- and cross-polarized components of the SHG signal via tuning the groove depth. Importantly, in this case \ce{Ga} ion FIB milling was proven to be non-destructive for the crystal structure: anisotropic features of the $\{111\}$-type surface were preserved after the fabrication.
In another example, Chen \emph{et~al.} explored a nanocavity composed of a monocrystalline \ce{Ag} nanocube on an \ce{Ag} crystalline flake, featuring double-resonance plasmonic modes and an ultrasmall gap, resulting in significantly enhanced SHG~\cite{Chen:2024c}. Here, the SHG from the $\{111\}$ surface of the \ce{Ag} flake is polarization-dependent, and its anisotropy can be controlled through the superposition of symmetries of the surfaces of the flake and the nanocube.

In many other examples of crystalline flake applications for nonlinear plasmonics, the anisotropic features of SHG were not used to engineer the response from plasmonic nanostructures, however they benefited from the improved fabrication quality, as in the case of linear nanoantennas (see Sec.~\ref{sec:PlasmonicNanoantennas}). 
For example, Celebrano \emph{et~al.} realized doubly-resonant, non-centrosymmetric monocrystalline \ce{Au} nanostructures that enabled highly efficient SHG~\cite{Celebrano:2015} and later used similar structures to study cascaded (SHG-mediated) third-harmonic generation~\cite{Celebrano:2019}. Wang \emph{et~al.} FIB-fabricated doubly resonant V-groove arrays in \ce{Ag} flakes to obtain more than three orders of magnitude enhancement of SHG as compared with bare silver sources (similar to Ref.~\cite{Boroviks:2025}, but omitting anisotropic SHG features).
In a study by Habibullah and Ishihara, SHG from metasurfaces consisting of square arrays of asymmetric structures were explored to demonstrate the control of the enhanced SHG~\cite{Habibullah:2022}. 

It is important to note here, that SHG enhancement can be achieved not only by virtue of plasmon resonances, but also by introducing additional asymmetries in the plasmonic nanostructures, which may be categorized into global asymmetry (antenna's overall shape) and local asymmetry (surface or antenna's gap morphology). 
In this connection, the SHG enhancement was studied in FIB-patterned \ce{Au} flakes: Meier \emph{et~al.} demonstrated that introducing geometric asymmetry to plasmonic antenna gaps significantly enhances SHG~\cite{Meier:2023}. 
While the global antenna mode increases the local field enhancement at the fundamental frequency, the local asymmetry is introduced at the position of the hotspot to circumvent the silencing effect. Intriguingly, even in the absence of the resonance at second-harmonic frequency, the overall SHG remains strong, despite being strongly damped by interband transitions in gold. 
In another work by Chen \emph{et~al.}, the symmetries of the modes propagating in a crystalline \ce{Au} two wire transmission line leveraged control over SHG intensity, directionality and polarization~\cite{Chen:2019a}.

\subsection{Nonlinear photoluminescence}

\begin{figure}[ht!]
    \centering
    \includegraphics[width=.8\linewidth]{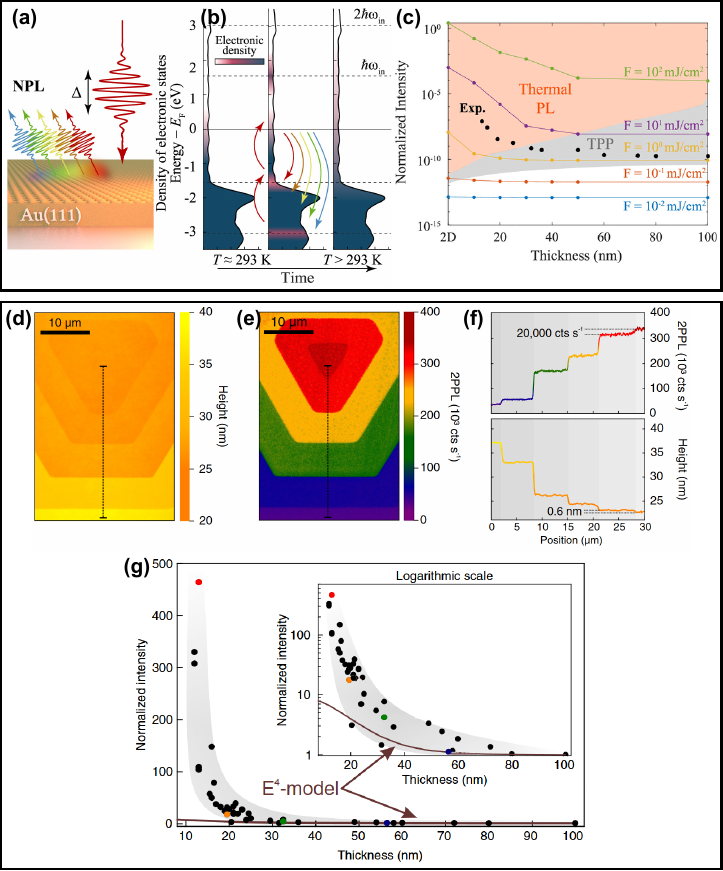}
    \caption{Nonlinear photoluminescence from \ce{Au} flakes. \pnl{a} Illustration of an \ce{Au} $\{111\}$-type illuminated by a femtosecond Gaussian pulse, along with the resulting nonlinear photoluminescence. \pnl{b} Dynamics of the occupied density of electronic states upon optical excitation. From the equilibrium state at room temperature, electrons are promoted to higher energy levels, and the associated holes are subsequently filled by emitting photons with different energies (colored curves). Once the pulse is gone, after a transient time of the order of tens of femtoseconds, the electronic states reach thermal equilibrium at a higher temperature than initially. \pnl{c} Nonlinear photoluminescence as a function of \ce{Au} film thickness and incident pulse fluence ($F$). Shaded areas indicate the regions where either thermal emission (orange) or two-photon photoluminescence (gray) dominates. The white area corresponds to the linear absorption regime. Black dots are experimental values taken from~\cite{Grossmann:2019}. \pnl{d} AFM image of a "terraced" region of \ce{Au} flake with thickness varying in the 20--40\,nm range and \pnl{e} two-photon photoluminescence image of the same flake region. \pnl{f} Thickness and photoluminescence profiles extracted from images in panels \pnl{d} and \pnl{e}. \pnl{g} Thickness dependence of the two-photon photoluminescence signal. The inset shows the same plot using a logarithmic scale. The brown line shows the expected intensity according to the $E^4$ law. \pnl{a--c} Reprinted with permission from Ref.~\cite{RodriguezEcharri:2023} and \pnl{d--g} from Ref.~\cite{Grossmann:2019}.
    \label{fig:NPL}}
\end{figure}

Nonlinear photoluminescence, also often referred to as multi-photon luminescence, is a radiative recombination process triggered by a nonlinear absorption [see schematic illustration in Fig.~\ref{fig:NPL}\pnl{a}]. 
In contrast to SHG, nonlinear absorption corresponds to a third- (or higher odd-) order nonlinear and non-parametric optical process, which does not impose any symmetry requirements on the medium. Thus, NPL may occur also in the bulk of centrosymmetric noble metal crystals, however in practice it also renders as a surface-like effect. Like in the case with quadrupolar (and higher-order) bulk SHG, the limiting factor is high reflectivity of gold in the visible and infrared wavelength range, which implies that the interaction takes place only in a near-interface region of approximately 10\,nm thickness, which corresponds to the penetration depth of the incident wave inside metal. 

Within the wave optics framework, nonlinear absorption is described by the imaginary part of the corresponding nonlinear susceptibility tensor ($\bm{\chi}^{(3)}$, $\bm{\chi}^{(5)}$, etc.).
Despite the fact that the measured value of the imaginary part of the third-order susceptibility of gold is comparatively high~\cite{Boyd:2014}, overall, two-photon absorption remains a weak process, which is limited by the small light-matter interaction volume, and requires extremely strong excitation fields to be observed (in an experimental setting it can be directly measured via so-called $z$-scan technique~\cite{Smith:1999,Goetz:2016}).

In a simplified quantum-mechanical picture, third-order nonlinear absorption corresponds to a cascaded of two-photon absorption, shown in Fig.~\ref{fig:NPL}\pnl{b}: upon absorption of a first photon, an electron in a conduction band gets promoted above the Fermi level, thereby creating a low-energy hole. As this event corresponds to an intraband transition within the $sp$-conduction band, it requires a source of additional momentum, which can be provided by subwavelength confinement of the electric field (e.g., surface discontinuities that produce plasmonic hotspots~\cite{Beversluis:2003}). In particular, this explains why, in the spectra presented in Fig.~\ref{fig:SHG}, the NPL signal from the flat $\{111\}$ crystalline surface [Fig.~\ref{fig:SHG}\pnl{d}] is considerably weaker than that from the polycrystalline surface [Fig.~\ref{fig:SHG}\pnl{c}]. Then, the second photon fills the hole in an $sp$-conduction band with an electron from a $d$-band, creating a new hole with even lower energy. This $sp$ hole has a lifetime of $\sim 1$\,ps, which implies that, when the pulse duration exceeds 1\,ps, the two-photon absorption efficiency decreases drastically with increasing pulse length~\cite{Biagioni:2009}.
Following some thermal relaxation process, the $d$-band hole and an electron from $sp$ band recombine by emitting a photon. As a result, a part of the the emitted photons have higher energy than the one of excitation, opposite to conventional linear photoluminescence characterized by emission at lower energies. Due to the random nature of the thermal relaxation process, both linear and nonlinear emission spectra are broad. In case of very strong excitation fields, the cascaded absorption can include absorption of more than two photons (thus the process is termed \emph{multi-photon}, and corresponds to higher-order odd nonlinear effect), which enlarges the energy gap between the recombining $d$-band hole and electron pushed above the Fermi level, thereby broadening the spectrum further into the wavelength~\cite{Biagioni:2012, Mejard:2016}. Here, we have provided an intuitive picture of nonlinear photoluminescence; however, for a more rigorous treatment, one should consider not individual electrons and holes, but rather their distribution across the electronic bands and the resulting complex population dynamics~\cite{RodriguezEcharri:2023}.

Although nonlinear photoluminescence from thin metal films was first experimentally discovered by Chen \emph{et~al.} already in 1981~\cite{Chen:1981}, and a few years later Boyd \emph{et~al.} performed a comprehensive study of nonlinear luminescence of various noble metals~\cite{Boyd:1986}, followed by a theoretical work by Apell \emph{et~al.}~\cite{Apell:1988}, the exact nature of this processes remains debated. 
Whilst, the previously described general mechanism appears to form consensus. the subtleties of the process are further complicated in case of nanostructured metals which support plasmonic resonances~\cite{Beversluis:2003, Mejard:2016, Malchow:2021, DellOva:2023, Sivan:2023, LoirettePelous:2024} or in very thin metal films which exhibit thickness-dependent mesoscopic effects~\cite{RodriguezEcharri:2021a, RodriguezEcharri:2021b, RodriguezEcharri:2023, Bowman:2024}. The latter was demonstrated with particular clarity for \ce{Au} flakes in experiment by Gro{\ss}man \emph{et~al.}~\cite{Grossmann:2019}, the main results of which are shown in Fig.~\ref{fig:NPL}\pnl{d--g}. As can be seen, for flake thicknesses below 20\,nm, the two-photon luminescence intensity deviates from the $E^4$-rule that is expected from classical theory. Originally, this unconventional behavior was attributed to the quantum confinement effects and modification of the gold band structure in thin crystalline films. However in recent works, by Rodr{\'i}guez~Echarri \emph{et~al.} and Sivan~\emph{et~al.} it was shown that this effect can be explained by extremely high electronic temperatures caused by increased absorption in thin gold films [see Fig.~\ref{fig:NPL}\pnl{c}]~\cite{RodriguezEcharri:2023, Sivan:2023}.

NPL microscopy, spectroscopy, and pump-probe measurements serve as versatile tools for exploring mesoscopic and ultrafast phenomena across condensed matter physics. Using crystalline metal flakes and nanostructures made of them allow to perform particularly clean experiments, for example studying ultrafast hot-carrier dynamics in \ce{Au}~\cite{Karaman:2024}. NPL microscopy of crystalline metal nanostructures has proven to be a powerful technique for characterization of plasmonic field enhancement with subwavelength resolution~\cite{Huang:2010a,Viarbitskaya:2013,Viarbitskaya:2013a,Fedou:2013,Knittel:2015,Cuche:2015,Cuche:2017,Podbiel:2017,Feichtner:2017}.

Finally, NPL from crystalline metal nanostructures also has a number of potential applications in nonlinear plasmonics. For example, Chen \emph{et~al.} revealed nanoantenna-mediated near-to-far-field energy coupling by exploring the modulation of transverse, antibonding, and higher-order longitudinal modes on \ce{Au} plasmonic nanoantennas via two-photon photoluminescence spectroscopy~\cite{Chen:2014}.
Dell’Ova \emph{et~al.} investigated the control of spatially distributed NPL in micrometer-scale cavities defined in crystalline \ce{Au} flakes, demonstrating that NPL can be generated at positions remote from the excitation spot and can be actively manipulated either by adjusting the polarization of the pulsed laser or through coherent two-beam excitation to modulate plasmon interference patterns~\cite{DellOva:2022}. Francescantonio \emph{et~al.} demonstrated coherent control over NPL from a single plasmonic nanoantennas by dual-beam pumping~\cite{Francescantonio:2022}. NPL signal from azimuthally chirped gratings was suggested to leverage sensing applications~\cite{Barman:2022,Chakraborty:2023}. Finally in terms of optical computing, \ce{Au} flakes were proposed to be used as building blocks for modal logic gates in nanoplasmonic circuitry~\cite{Viarbitskaya:2013,Kumar:2021}.

\section{Material platform for extreme polaritonics beyond classical electrodynamics}
\label{sec:quantum}

As mentioned in Sec.~\ref{Sec:devices}, crystalline metal flakes enable applications in quantum plasmonics, by interfacing quantum light sources with plasmonic cavities and waveguides. However, besides serving as high-quality material platform for prototypical quantum plasmonic devices, unstructured flakes themselves support non-classical electronic and optical responses.
As the electronic band structure of polycrystalline nanostructures is likely to differ from that of monocrystals~\cite{Gleiter:2001}, due to intrinsic randomness of the former and perfect order of the latter, experimental observation of subtle mesoscopic effects using polycrystalline samples is often obscured. Hence, owing to their well-defined atomic structure, the crystalline flakes render as a "sandbox" for sophisticated experiments in solid state physics.
In this section we review some examples of exotic phenomena in solid state physics and extreme states of light.

\subsection{Nonlocal effects}

The advancing ability to synthesize and nanofabricate metallic nanostructures -- together with novel experimental techniques for genuine nanoscale exploration -- has renewed theoretical interests in the nonlocal electrodynamics of plasmons in metals~\cite{Monticone:2025}. In short, nonlocal electrodynamics refers to a description of how materials respond to electromagnetic fields in which the response at a given point depends not only on the field at that point but also on the fields in its immediate surroundings~\cite{Monticone:2025}. In Fourier space, this implies that the response function -- in practice the dielectric permittivity $\varepsilon$ -- now depends on both the frequency $\omega$ and the wave vector $\mathbf{k}$ of the light, i.e., $\varepsilon(\omega,\mathbf{k})$. The nonlocal polarizability of the quantum electron gas in a metal originates from its finite compressibility. In contrast, within the local-response approximation, the electron gas is assumed to be incompressible and is therefore displaced as a rigid body in response to an electric field, whereas in reality it can both be displaced and deformed~\cite{Mortensen:2021a}. Quantum nonlocal effects can be accounted for in the classical electrodynamics at various levels of response formalism, ranging from semi-classical nonlocal hydrodynamics~\cite{Raza:2015,Ciraci:2013} to surface-response formalism~\cite{Feibelman:1982, Mortensen:2021a,Mortensen:2021b}.

In plasmonics, nonlocal effects emerge at the nanoscale, where the size of structures is comparable to the electron mean-free path and where mesoscopic electrodynamic effects manifest at the surfaces~\cite{Mortensen:2021a,Mortensen:2021b}. Here, the classical local assumption -- that the material’s polarization depends only on the local electric field -- is challenged, and quantum and spatial dispersion effects must be considered. This leads to shifts in plasmon resonances, modified field confinement, and altered optical responses. This is particularly important in geometries with abruptly changing surface morphologies, which would otherwise exhibit singular or diverging responses within the local-response approximation.

There has been particular interest in size-dependent spectral resonance shifts and spectral broadening in metallic nanoparticles and nanostructures. These effects are often attributed to the nonlocal response of the electrons (arising from the finite compressibility of the quantum electron gas) and to possible quantum spill-out (linked to the finite work function confining the electrons). The phenomenon of size-dependent broadening has attracted long-standing attention, though its quantitative interpretation is challenging due to the coexistence of nonlocal effects with more classical Ohmic losses, the latter potentially overshadowing the former in real experiments. As an example, Hajisalem \emph{et~al.}~\cite{Hajisalem:2014} concluded that in their particle-on-mirror \ce{Au} structures, surface roughness, rather than nonlocal optical effects, is responsible for saturation of plasmonic resonance shifts in at sub-nanometer spacer thicknesses. In this context, crystalline flakes are particularly attractive, as their lower bulk Ohmic losses and surface-scattering losses may help disentangle these contributions and enable an unambiguous connection between size-dependent plasmonic corrections and the nonlocal electrodynamics of the electron system in metals. Turning from localized to propagating plasmons, the rationale for favoring crystalline flakes over amorphous films is similar: nonlocal shifts in the complex dispersion relation must be distinguished from spectral broadening caused by bulk damping. Indeed, this was the underlying motivation for the gap-plasmon experiments of Boroviks \emph{et~al.}~\cite{Boroviks:2022}, where undesirable background Ohmic and scattering losses were minimized by using \ce{Au} flakes instead of amorphous films and high-quality \ce{Al2O3} layers grown by ALD rather than by CVD.

The crystalline quality of \ce{Au} flakes thus appears a crucial prerequisite for observing quantum nonlocal corrections to the classical electrodynamics of plasmonic nanostructures~\cite{Mortensen:2021a}.

\subsection{Enabling quantum effects and exotic phenomena}

An overarching hypothesis and motivation for exploiting crystalline metal flakes in the exploration of possible quantum effects in plasmonics is that -- as for the nonlocal corrections -- such effects could in general be compromised or obscured by unintended loss mechanisms associated with metals. We hypothesize that employing high-quality crystalline flakes, which are expected to exhibit reduced plasmonic losses, provides a promising pathway toward minimizing extrinsic damping and thereby enabling clearer observation and possible control of genuine quantum plasmonic phenomena. However, it remains an open question whether the reduction in losses would be sufficient to clearly disentangle extrinsic damping from intrinsic quantum effects.

Bowman \emph{et~al.} demonstrates how thin monocrystalline \ce{Au} flakes enable the observation of quantum-mechanical effects in linear photoluminescence, providing unprecedented insight into hot-carrier dynamics in plasmonic nanostructures~\cite{Bowman:2024}. Unlike polycrystalline or bulk metals, the atomically ordered structure of the flakes minimizes extrinsic scattering and defects, allowing subtle luminescence signals -- arising from electron-hole recombination under interband excitation -- to be clearly detected and analyzed. Experiments, supported by first-principles simulations, reveal that quantum effects in luminescence become pronounced as the flake thickness decreases, observable even in flakes up to 40\,nm, due to the out-of-plane discretization of the electronic band structure near the Fermi level.

Another motivation for using crystalline metallic flakes is the ability to realize highly well-defined geometries, including the integration of metallic leads with plasmonic resonators. As an example of this, a recent work by Zurak \emph{et~al.}, which investigated how electrical charging modulates the optical response of single plasmonic nanoresonators, revealing nonclassical surface effects beyond the scope of standard classical models. By combining lock-in detection techniques with direct electrical gating, the authors measured changes in light scattering that could be quantitatively interpreted using surface-response functions, capturing the quantum surface effects associated with even minute charging at the surface region of the resonator~\cite{Zurak:2024}.

Zheng \emph{et~al.} demonstrated electrical control of light-matter interactions at the atomic scale by integrating semiconductor monolayers into nanometer-gap crystalline nanocube-on-mirror plasmonic cavities~\cite{Zheng:2025}. Here, the ultrastrong fields in these low-loss, high-quality crystalline cavities enable reversible modulation of Rabi splitting when excitons couple to the cavity. The crystalline nature of the metallic cavities is essential, providing the field confinement, low loss, and reproducibility needed to achieve strong coupling, precise electrical control, and efficient emission from excitons in 2D materials.

Spektor \emph{et~al.} performed sophisticated experiments with structured light beams using time-resolved photoemission electron microscopy (PEEM) technique, demonstrating that the optical angular momentum modes of light can be shrunk down to the nanometer scale on a metallic surface~\cite{Spektor:2017}. In this kind of experiment, a metal surface of exceptional smoothness is required to achieve strong static electric fields for photoelectron extraction. Furthermore, crystalline flakes have also been a key enabling ingredient in explorations of Skyrmion dynamics associated with the vector fields of SPPs, where achieving high spatial and temporal resolution critically depends on minimizing plasmonic scattering and losses~\cite{Davis:2020,Dreher:2024a,Schwab:2025,Tsesses:2025}.
Another area of fundamental research revitalized by the access to well-defined metallic surfaces is the study of Casimir--Lifshitz forces~\cite{Munkhbat:2021, Schmidt:2023, Kucukoz:2024, Hoskova:2025}. In this context, the intrinsically weak quantum forces arising from vacuum fluctuations can be readily obscured by unintended surface roughness and imperfections in topography, while plasmonic losses further influence the interactions.

By employing transient absorption spectroscopy methods, Wang \emph{et~al.} showed evidence of strong vibrational coupling between two crystalline plasmonic resonators at room temperature~\cite{Wang:2019}. In another study of coupled resonators, Hensen \emph{et~al.} further revealed that even within a single plasmonic mode, local field dynamics can exhibit significant spatial variations in apparent $Q$-factor and resonance frequency, arising from crosstalk between adjacent modes, as observed via time-resolved photoemission electron microscopy on femtosecond timescales~\cite{Hensen:2019}. Finally, Schurr \emph{et~al.}, demonstrated that milling one-dimensional array of nanoslits with alternating coupling strength in an \ce{Au} flake allows to observe topologically protected modes in Su--Schrieffer--Heeger (SSH) chains~\cite{Schurr:2025}. 

So far, we have considered phenomena that are implicitly linked to the quantum properties of matter, namely the quantum-mechanical aspects of the electron gas supporting the plasmonic response. However, another intriguing aspect of quantum plasmonics concerns the quantum nature of the light fields interacting with plasmonic nanostructures~\cite{Bozhevolnyi:2017a}. As an example of such quantum optical effects dealing with single plasmon quanta, Pres \emph{et~al.} demonstrated the local detection of nanoscale plasmon quantum wave packets using plasmon-assisted electron emission in coherent 2D nanoscopy~\cite{Pres:2023}. Notably, the observation of a quantum coherence oscillating at the third harmonic of the plasmon frequency directly revealed the superposition of non-adjacent plasmon occupation states, providing a fingerprint of the plasmon quantum wave packet. The use of high-quality crystalline metal flakes was crucial for these explorations: their atomically smooth, defect-free surfaces minimize plasmon damping and loss, enabling reproducible, ultrastrong local fields that make the detection of quantum wave packet dynamics possible.
In a similar spirit, Dreher \emph{et~al.} also exploited crystalline metallic flakes, harvesting their atomically smooth, defect-free surfaces to support reproducible, ultrastrong SPP fields with minimal damping, enabling momentum space separation of quantum path interferences between photons and SPPs in nonlinear photoemission microscopy~\cite{Dreher:2024}.

\subsection{Tamm--Shockley surface states and 2D plasmons}

Beyond their bulk properties, crystalline metals are increasingly relevant for surface phenomena too. In particular, surface science has long established~\cite{Inglesfield:1982} that noble-metal $\{111\}$ surfaces support in-plane conductive surface states, known as Tamm--Shockley (TS) states~\cite{Tamm:1932,Shockley:1939}. These states exhibit a parabolic, free-electron-like dispersion and host a two-dimensional electron gas confined near the surface. The TS surface state is characterized by a surface conductivity $\sigma_\parallel(\omega,\mathbf{k}_\parallel)$, which depends on both frequency $\omega$ and in-plane wavevector $\mathbf{k}_\parallel$. Its influence on the electromagnetic response can be incorporated into classical electrodynamics by modifying the boundary conditions to include the surface conductivity, providing a direct connection to the Feibelman surface-response function, where $d_\parallel \propto \sigma_\parallel$~\cite{Feibelman:1982}.
Crystalline noble-metal nanoparticles typically feature multiple facets corresponding to the crystal planes of the material~\cite{Quan:2013}, giving rise to morphology- and polarization-dependent plasmonic resonances~\cite{Myroshnychenko:2008a,Yoon:2019,Elliott:2022,RodriguezEcharri:2021a}. In contrast, large planar flakes are dominated by $\{111\}$ facets, although their edges can expose other crystal planes~\cite{Boroviks:2018}.

The electrodynamic response of a homogeneous 2D electron gas, potentially embedded in a general dielectric environment, and the associated plasmon excitations -- characterized by a square-root dependence $\omega\propto \sqrt{k_\parallel}$ of plasmon energy on wave vector~\cite{Ando:1982} -- have been extensively studied theoretically. However, when turning to the TS surface states, the associated 2D plasmons are screened by the electronic states of the bulk that supports the surface, which changes the square-root dispersion into a linear acoustical dispersion $\omega\propto k_\parallel$~\cite{Pitarke:2007}. Following early general work by Nagao \emph{et~al.} on the dispersion and damping of a 2D plasmon in a metallic surface-state band~\cite{Nagao:2001} and work by Silkin \emph{et~al.} on these acoustic surface plasmons in the noble metals \ce{Cu}, \ce{Ag}, and \ce{Au}~\cite{Silkin:2005}. More recently, Yan \emph{et~al.} have interestingly shown how the TS surface states can be reinterpreted as topologically derived surface states of a topological insulator~\cite{Yan:2015c}. 

Experimentally, Mugarza \emph{et~al.} have used angle-resolved photoemission to explore electron confinement in surface states on a stepped \ce{Au} surface~\cite{Mugarza:2001}, while TS states have also been probed with synchrotron radiation photoemission spectroscopy~\cite{Silva:2007}.

The low-energy acoustic plasmons are tightly confined to the surface with exceeding high $\mathbf{k
}_\parallel$, making their optical excitation challenging. However, this can be overcome using angle-resolved electron energy loss spectroscopy, as demonstrated first for acoustic plasmons in the \{001\} surface of beryllium (\ce{Be})~\cite{Diaconescu:2007} and later also for \ce{Au} $\{111\}$~\cite{Park:2010}, with the acoustic plasmons maintaining their linear acoustic dispersion even in the presence of the bulk response of the crystal and its single-particle excitations~\cite{Vattuone:2013}. 

In a more recent study, Dreher \emph{et~al.} investigated quantum coherent coupling between nano-focused surface plasmons and electronic states in solids, demonstrating a pathway toward Floquet engineering at the nanoscale~\cite{Dreher:2023}. Using \ce{Au} $\{111\}$ TS surface states, the experiment revealed above-threshold electron emission via absorption of up to seven SPP quanta. Time-resolved photoelectron spectroscopy reveals that the SPP-surface state interaction preserves quantum coherence during emission, analogous to light-driven above-threshold photoemission. These results establish a platform for coherent, nanoscale light-matter interactions and potentially suggest that SPP-based Floquet engineering in nano-optical systems could be practically realized.

\subsection{Enabling ideal mirrors in 2D material heterostructures}
\label{sec:IdealMirrors}

Atomically smooth crystalline metal flakes have emerged as nearly ideal optical mirrors in the IR and mid‑IR spectral ranges, providing surfaces free from the scattering and loss channels typical of polycrystalline films. Their atomically flat structure ensures minimal surface roughness and the absence of grain boundaries, which allows for coherent reflection of electromagnetic waves over long distances. This property is particularly advantageous in van~der~Waals (vdW) heterostructures and polaritonic systems, where placing a 2D material -- such as a crystal supporting phonon or plasmon polaritons -- near the mirror leads to strong light-matter interactions, enhanced confinement, and long-range polariton propagation. 
This flatness is also of particular importance for various scanning probe microscopies. Already in 2006, Dahanayaka \emph{et~al.} demonstrated the advantages of using thin \ce{Au} flakes as optically semi-transparent with atomic flatness for probing self-assembled monolayers using scanning tunneling microscopy~\cite{Dahanayaka:2006}. Later, this platform was proven to be highly beneficial in TERS~\cite{Deckert-Gaudig:2009a, Pienpinijtham:2012, Pashaee:2013, Pashaee:2015, Richard:2020}. In recent work by Sabzehparvar \emph{et~al.}, scanning electrochemical imaging of a photo-oxidation reaction on thin \ce{Au} flakes further illustrates the advantages of this platform: monocrystalline , thin-walled electrodes provide high spatial resolution, enhanced chemical sensitivity, and stable performance over extended measurements, enabling nanoscale electrochemical studies that were previously limited by electrode fragility and fabrication constraints~\cite{Sabzehparvar:2026}.

In near-field optical probes such as scanning near-field optical microscopy (SNOM), the pristine surfaces of crystalline flakes preserve the momentum and coherence of polaritons, resulting in sharp interference patterns and fringes and consequently reliable and quantitative mapping of polaritonic dispersion~\cite{Menabde:2022a,Menabde:2022b,Menabde:2022c,Casses:2024,Heiden:2025}. Moreover, these high-quality mirrors play a key role in cavity QED with 2D excitons and in nanoparticle-on-mirror plasmonic nanocavities, where atomic-scale flatness maximizes field enhancement and minimizes losses~\cite{Liu:2022,Wang:2023a}. Together, these attributes make crystalline metal flakes indispensable components for next-generation IR/mid‑IR photonic and quantum devices.

\section{Outlook and future directions}

Noble metals have been the subject of persisting commercial, technological, and scientific interest for centuries. The history of gold mining dates back to ancient civilizations, and as of 2025, it was estimated that overall more than $2\times 10^5$\,tonnes
of gold has been mined throughout human history~\cite{Gold:2026}. The burst of micro- and nanotechnology in the second half of the 20th century has further increased this interest. Although only 7\% of the worldwide annual gold demand is concerned with the technological demands, gold and other noble metals remain crucial materials for the modern electronics industry. Beyond their role in conventional electronics, noble metals have unique optical properties, which have underpinned the development of the field of plasmonics. 

While the promises of plasmonics in its early days were burgeoning~\cite{Barnes:2003, Ozbay:2006, Atwater:2007} -- including subwavelength optical waveguides, interconnects and ultrafast modulators, negative-index metamaterials and other emerging applications -- efficient implementation and commercialization of these devices have proven to be challenging. To a large extent, this difficulty is due to the inevitable absorption losses in metals at optical frequencies. At the time of writing this review, the research community has reached a consensus that the initial promises of plasmonics have proven elusive, although in some cases losses in plasmonic systems can be exploited~\cite{Boriskina:2017, Khurgin:2015c}. Hence, plasmonics has found its niche applications, some of which (for example, surface plasmon resonance-based refractive index sensing~\cite{Jodaylami:2025} and SERS~\cite{Langer:2020}) have been successfully commercialized. Furthermore, noble metal nanoparticles and nanomaterials are widely studied for emerging applications in medicine and biology, both for clinical treatment and sensing~\cite{Webb:2014,Yang:2015,Sasidharan:2015}. In some cases, gold nanoparticles are already widely used; for example, one of the most widespread applications of gold nanoparticles is biolabeling in rapid antigen tests. 

Yet, the objective of this review is to direct the attention of the research community toward the potential to at least partly mitigate plasmonic losses by employing noble metals in their nearly perfect crystalline form, thereby minimizing issues associated with structural imperfections in conventional polycrystalline films.
However, we should refrain from excessively optimistic vision: crystalline flakes are not a panacea for all challenges in plasmonics, ass Ohmic losses are inevitable even in the finest form of noble metals~\cite{Khurgin:2015c}.
Nevertheless, this review highlights several areas where the flakes do make a noticeable difference. As outlined in Sec.~\ref{sec:NanoMicroFab}, when patterned using state-of-the-art nanofabrication tools (\ce{He} FIB or wet/dry etching following EBL), the resulting crystalline nanostructures exhibit exceptional surface quality and minimal deviations from the nominal dimensions and target morphologies, i.e., extremely tight tolerances.
Sec.~\ref{Sec:devices} presents a plethora of examples in which monocrystalline flake-based nanoantennas and devices outperform their polycrystalline counterparts.
Quantum plasmonic applications are especially compelling, as they enable the realization of very bright quantum light sources~\cite{Bozhevolnyi:2017b}. As discussed in Sec.~\ref{Sec:nonlinear}, the surfaces of crystalline metals exhibit a stronger nonlinear optical response that is tunable and stable. Finally, monocrystalline metal flakes present a sandbox material for fundamental experimental research in solid-state physics. Even in their bare form, crystalline metal surfaces exhibit exotic phenomena, and when used as a supporting material platform for 2D material heterostructures, they allow to perform clean experiments in mesoscopic and extreme light-confinement regimes (see examples in Sec.~\ref{sec:quantum}.

We would like to conclude this review with a parting (and, possibly, wishful) thought, expressed by Prof. Gary~Leach at the Gold 2025 conference in San Sebastian, Spain (paraphrased here)~\cite{Leach:2025}: the success of the semiconductor electronics industry owes a great deal to the material research community, which developed inexpensive, large-scale, and reliable methods for the growth of \ce{Si} monocrystals. In plasmonics, prototypical devices based on monocrystalline metals have already demonstrated substantial improvements over their conventionally fabricated counterparts. Perhaps investment in materials science and development of fabrication and patterning methods for the high-quality, wafer-level processing of monocrystalline noble metals would revitalize the field of plasmonics, and open new opportunities for applications, and further drive fundamental research.

\section*{Funding}

N.~A.~M. acknowledges support from DNRF -- Danish National Research Foundation (project No.~DNRF147).

J.-S. H. acknowledges the support from DFG -- Deutsche Forschungsgemeinschaft (project No.~437527638 \& 398816777).

\section*{Acknowledgments}

\section*{Disclosures}

The author declares no conflicts of interest.

\bibliography{references}

\newpage
\renewcommand\thepage{S\arabic{page}}
\setcounter{page}{1}
\setcounter{table}{0}
\renewcommand{\thetable}{S\arabic{table}}

\title{Supporting information ---\newline Crystalline metal flakes: Platforms for advanced plasmonics and hybrid 2D material architectures}

\author{Sergejs Boroviks\,\orcidlink{0000-0002-3068-0284}\authormark{1}}

\author{Siarhei Zavatski\,\orcidlink{0000-0003-4530-4545}\authormark{1}}

\author{Thorsten Feichtner\,\orcidlink{0000-0002-0605-6481}\authormark{2}}

\author{Jer-Shing Huang\,\orcidlink{0000-0002-7027-3042}\authormark{3}}

\author{Olivier J. F. Martin\,\orcidlink{0000-0002-9574-3119}\authormark{1}}

\author{Bert Hecht\,\orcidlink{0000-0002-4883-8676}\authormark{2}}

\author{N.~Asger~Mortensen\,\orcidlink{0000-0001-7936-6264}\authormark{4,5,*}}

\address{\authormark{1}Nanophotonics and Metrology Laboratory, Swiss Federal Institute of Technology Lausanne (EPFL), EPFL-STI-NAM, Station 11, Lausanne, CH-1015 Switzerland}
\address{\authormark{2}Nano-Optics and Biophotonics Group, Experimentelle Physik 5, Physikalisches Institut, Universität Würzburg, Am Hubland, Würzburg, Germany}
\address{\authormark{3}Leibniz Institute of Photonic Technology, Albert-Einstein-Str. 9, 07745 Jena, Germany}
\address{\authormark{4}POLIMA---Center for Polariton-driven Light--Matter Interactions, University of Southern Denmark, Campusvej 55, DK-5230 Odense M, Denmark} 
\address{\authormark{5}D-IAS---Danish Institute for Advanced Study, University of Southern Denmark, Campusvej 55, DK-5230 Odense M, Denmark} 
\email{\authormark{*}namo@mci.sdu.dk}

\newpage
\begingroup
\setlength{\LTleft}{-2.5cm}  % Shift left margin
\setlength{\LTright}{-2.5cm} % Shift right margin
{
%\small
%\input{Table_ChemSynthesisExtended}}
\tiny
\begin{longtable}{|l|l|p{3.5cm}|p{1.5cm}|p{2cm}|p{3cm}|l|l|}
    \caption{Overview of the \ce{Au} flake chemical synthesis methods in chronological order\label{tab:SI}}\\
    \hline
Year & Reference \& doi  & Method  & Solvent  & Additional chemicals & Feature   & \makecell{Typical lateral\\ size/area} & \makecell{Typical\\ thickness}  \\ \hline
    \endfirsthead

    \hline
Year & Reference  \& doi  & Method  & Solvent  & Additional chemicals & Feature   & \makecell{Typical lateral\\ size/area} & \makecell{Typical\\ thickness}  \\ \hline
    \endhead

    \hline
    \endfoot

    \hline
    \endlastfoot
1994 & \makecell[l]{\cite{Brust:1994} \\  \href{https://doi.org/10.1039/C39940000801}{10.1039/C39940000801}} & Reduction by sodium borohydride & Water-toluene & alkanethiol & First report of plate-like particles via reduction of \ce{AuCl4-} in a   two-phase system & $2-\SI{2.5}{\nano\meter}$ &  \\
2004 & \makecell[l]{\cite{Shankar:2005} \\   \href{https://doi.org/10.1021/cm048292g}{10.1021/cm048292g}} & Reduction by lemongrass extract & Water &  &  & $\sim\SI{2}{\micro\meter}$ &  \\
2004 & \makecell[l]{\cite{Shankar:2004} \\   \href{https://doi.org/10.1038/nmat1152}{10.1038/nmat1152}} & Reduction by lemongrass extract & Water &  & Biological synthesis & $\sim\SI{2}{\micro\meter}$ &  \\
2004 & \makecell[l]{\cite{Sun:2004} \\    \href{https://doi.org/10.1002/ange.200461013}{10.1002/ange.200461013}} & Reduction using ortho-phenylenediamine at room temperature & Water &  & Large scale by a mild wet-chemical route & $\sim\SI{2}{\micro\meter}$ & $\sim\SI{50}{\nano\meter}$ \\
2005 & \makecell[l]{\cite{Li:2005} \\    \href{https://doi.org/10.1021/jp0520998}{10.1021/jp0520998}} & Microwave heating of ionic liquid & ioniq liquid 1-Butyl-3-methylimidazolium tetrafluoroborate &  &  & $<\SI{30}{\micro\meter}$ & $\sim\SI{50}{\nano\meter}$ \\
2006 & \makecell[l]{\cite{Li:2006} \\    \href{https://doi.org/10.1002/adfm.200500209}{10.1002/adfm.200500209}} & Polyol & Ethylene glycol & Polyvinylpyrrolidone & Size and shape control via concentrations and temperature & $2-\SI{20}{\micro\meter}$ & $70-\SI{600}{\nano\meter}$\\
2006 & \makecell[l]{\cite{Guo:2006} \\   \href{https://doi.org/10.1016/j.colsurfa.2005.11.075}{10.1016/j.colsurfa.2005.11.075}} & Polyol at \SI{95}{\degreeCelsius} & Ethylene glycol & Aniline &  & $<\SI{15}{\micro\meter}$ & $\sim\SI{5}{\nano\meter}$\\
2006 & \makecell[l]{\cite{Kan:2006} \\   \href{https://doi.org/10.1021/jp054800d}{10.1021/jp054800d}} & Polyol at \SI{150}{\degreeCelsius} & Ethylene glycol & Polyvinylpyrrolidone &  &  $<\SI{50}{\micro\meter}$ & $\sim\SI{70}{\nano\meter}$\\
2007 & \makecell[l]{\cite{Kawasaki:2007} \\    \href{https://doi.org/10.1246/cl.2007.1038}{10.1246/cl.2007.1038}} & Thermolysis in two-component ionic liquid at \SI{220}{\degreeCelsius} & & ce{C4ImPF6} and \ce{C8ImPF6} & First report o f millimeter scale flakes & $50-\SI{900}{\micro\meter}$ & $10-\SI{50}{\nano\meter}$\\
2008 & \makecell[l]{\cite{Lim:2008} \\   \href{https://doi.org/10.1021/la801803z}{10.1021/la801803z}} & Reduction with polyvinylpyrrolidone at \SI{100}{\degreeCelsius} & Water &  & No caping agents & $<\SI{30}{\micro\meter}$ &  \\
2010 & \makecell[l]{\cite{Radha:2010} \\   \href{https://doi.org/10.1007/s12274-010-0040-6}{10.1007/s12274-010-0040-6}} & Thermolysis at \SI{130}{\degreeCelsius} & Toluene & tetraoctylammonium bromide & Development of the Brust-Schiffrin method, on substrate growth & $>\SI{100}{\micro\meter}$ & $<\SI{1}{\micro\meter}$ \\
2011 & \makecell[l]{\cite{Radha:2011} \\   \href{https://doi.org/10.1021/cg1015548}{10.1021/cg1015548}} & Thermolysis at \SI{130}{\degreeCelsius} & Toluene & tetraoctylammonium bromide & Development of the Brust-Schiffrin method, on substrate growth & $>\SI{100}{\micro\meter}$ & $<\SI{1}{\micro\meter}$ \\
2012 & \makecell[l]{\cite{Radha:2012} \\ } & Thermolysis at 130℃ & Toluene & tetraoctylammonium bromide & Development of the Brust-Schiffrin method, on substrate growth & $>\SI{100}{\micro\meter}$ & $<\SI{1}{\micro\meter}$ \\
2013 & \makecell[l]{\cite{Qin:2013} \\   \href{https://doi.org/10.1021/ja406107u}{10.1021/ja406107u}} & 2D template-directed synthesis in a iridescent solution composed of   lamellar membranes & Water & dodecylglyceryl itaconate & Tunable thickness & $>\SI{10}{\micro\meter}$ & $5-\SI{40}{\nano\meter}$ \\
2014 & \makecell[l]{\cite{Niu:2014} \\ \href{https://doi.org/10.1038/ncomms4313}{10.1038/ncomms4313}} & Synthesis in a polymer-free lamellar hydrogel composed of   self-assembled  membranes & Water & hexadecylglyceryl maleate & Ultrathin gold layers & $>\SI{10}{\micro\meter}$ &$>\SI{3.6}{\nano\meter}$ \\
2015 & \makecell[l]{\cite{Park:2015} \\ \href{https://doi.org/10.1021/acs.nanolett.5b01677}{10.1021/acs.nanolett.5b01677}} & Electron beam-induced growth & Water &  &  & $0.2-\SI{1}{\micro\meter}$ & $\sim\SI{50}{\nano\meter}$ \\
2015 & \makecell[l]{\cite{Wu:2015} \\   \href{https://doi.org/10.1002/crat.201400429}{10.1002/crat.201400429}} & Polyol at \SI{50}{\degreeCelsius} & Ethylene glycol & Aniline & On-substrate regrowth & $<\SI{50}{\micro\meter}$ & $>\SI{100}{\nano\meter}$ \\
2015 & \makecell[l]{\cite{Zhou:2015} \\   \href{https://doi.org/10.1002/adma.201405121}{10.1002/adma.201405121}} & Chemical reduction at \SI{60}{\degreeCelsius} & Water & $\beta$-lactoglobulin amyloid fibrils & & $<\SI{310}{\micro\meter}$ & $<\SI{800}{\nano\meter}$ \\
2015 & \makecell[l]{\cite{Zhou:2015a} \\   \href{https://doi.org/10.1039/C4CC10040A}{10.1039/C4CC10040A}} & Photochemical reduction in lamellar liquid crystals & Water & polyoxyethylene sorbitan monopalmitate and camphorsulfonic acid & Improved ontrol over thickness & $<\SI{30}{\micro\meter}$ & $>\SI{100}{\nano\meter}$ \\
2016 & \makecell[l]{\cite{Wang:2016} \\   \href{https://doi.org/10.1039/C6RA15909E}{10.1039/C6RA15909E}} & Pyrolysis in a tube furnace at \SI{500}{\degreeCelsius} &  &  &  & $2-\SI{3}{\micro\meter}$ &  $\sim\SI{50}{\nano\meter}$ \\
2016 & \makecell[l]{\cite{Fang:2016} \\   \href{https://doi.org/10.1016/j.msec.2016.03.113}{10.1016/j.msec.2016.03.113}} & Silk fibrils-assisted thermolysis & Water & \ce{NaOH} or \ce{HCl} &  & $\sim\SI{10}{\micro\meter}$ &  \\
2016 & \makecell[l]{\cite{Bhosale:2016} \\   \href{https://doi.org/10.1002/slct.201600044}{10.1002/slct.201600044}} & Sonochemical synthesis at room temperature & Water & dimethyl sulfoxide &  & $\sim\SI{5}{\micro\meter}$ & $\sim\SI{20}{\nano\meter}$ \\
2017 & \makecell[l]{\cite{Chen:2017} \\   \href{https://doi.org/10.1039/C7TB02792C}{10.1039/C7TB02792C}} & Chitin nanofibrils-assisted & Water &  &  & $<\SI{10}{\micro\meter}$ & $\sim\SI{70}{\nano\meter}$ \\
2018 & \makecell[l]{\cite{Lv:2018} \\   \href{https://doi.org/10.1021/acssuschemeng.8b02954}{10.1021/acssuschemeng.8b02954}} & Silk fibrils-assisted  at \SI{80}{\degreeCelsius} & Water & \ce{HCl} + \ce{NaCl}&  & $<\SI{1000}{\micro\meter}$ & $<\SI{800}{\nano\meter}$ \\
2018 & \makecell[l]{\cite{He:2018} \\   \href{https://doi.org/10.1039/C8NR04942D}{10.1039/C8NR04942D}} & Photochemical synthesis & Water & methylene blue &  & $<\SI{4}{\micro\meter}$ & $>\SI{2}{\nano\meter}$ \\
2018 & \makecell[l]{\cite{Krauss:2018} \\   \href{https://doi.org/10.1021/acs.cgd.7b00849}{10.1021/acs.cgd.7b00849}} & Polyol at $70-\SI{90}{\degreeCelsius}$ & Ethylene glycol &  & Improved ontrol over thickness via temperature & $<\SI{180}{\micro\meter}$ & $>\SI{20}{\nano\meter}$ \\
2019 & \makecell[l]{\cite{Golze:2019} \\   \href{https://doi.org/10.1021/acs.nanolett.9b02215}{10.1021/acs.nanolett.9b02215}} & Photochemical synthesis & Water & Polyvinylpyrrolidone, methanol & Plasmon-mediated synthesis of periodic arrays using substrate-immobilized   seeds & $<\SI{1}{\micro\meter}$ & $>\SI{20}{\nano\meter}$ \\
2019 & \makecell[l]{\cite{Yang:2019c} \\   \href{https://doi.org/10.3390/nano9040595}{10.3390/nano9040595}} & Vapor-phase synthesis from \ce{AuCl} at \SI{200}{\degreeCelsius} & - &  & Low-Temperature Vapor-Phase Synthesis & $<\SI{5}{\micro\meter}$ & $\sim\SI{100}{\nano\meter}$\\
2020 & \makecell[l]{\cite{Ye:2020} \\   \href{https://doi.org/10.1002/adfm.202003512}{10.1002/adfm.202003512}} & Room temperature reduction using methylene orange & Water & \ce{FeBr3} & Size and thickness control via adjusting concentration & $0.15-\SI{2}{\micro\meter}$ & $7-\SI{20}{\nano\meter}$ \\
2021 & \makecell[l]{\cite{Farkas:2021} \\   \href{https://doi.org/10.1021/acs.jpcc.1c08404}{10.1021/acs.jpcc.1c08404}} & Diffusion in agarose hydrogel using the Turkevich method at room   temperature & Water & \ce{C6H5O7.3Na.2H2O} & Size and shape control via reaction-diffusion & $>\SI{20}{\micro\meter}$ & $\sim\SI{50}{\nano\meter}$ \\
2022 & \makecell[l]{\cite{Kiani:2022} \\   \href{https://doi.org/10.1021/acs.chemmater.1c03908}{10.1021/acs.chemmater.1c03908}} & Polyol at \SI{90}{\degreeCelsius} & Ethylene glycol & Various halides & Gap-assisted synthesis & $<\SI{150}{\micro\meter}$ & $>\SI{20}{\nano\meter}$ \\
2022 & \makecell[l]{\cite{Ma:2022} \\   \href{https://doi.org/10.1021/acs.langmuir.2c02404}{10.1021/acs.langmuir.2c02404}} & Reduction with EDTA at \SI{150}{\degreeCelsius} & Water & \ce{NaOH}, ethylenediaminetetraacetic acid & Hydrothermal synthesis & $3-\SI{11}{\micro\meter}$ & $80-\SI{350}{\nano\meter}$\\
2024 & \makecell[l]{\cite{Wu:2024} \\   \href{https://doi.org/10.1021/acsami.4c15121}{10.1021/acsami.4c15121}} & Recrystallization of polycrystalline film & - & tetraoctylammonium bromide & Recrystallization & $<\SI{120}{\micro\meter}$ & $\sim\SI{30}{\nano\meter}$ \\
2024 & \makecell[l]{\cite{Pan:2024} \\   \href{https://doi.org/10.1038/s41467-024-47133-7}{10.1038/s41467-024-47133-7}} & Polyol at \SI{90}{\degreeCelsius} and subsequent thinning & Ethylene glycol & Cysteamine and chloroform & Subsequent thinning (etching) in cysteamine & $>\SI{100}{\micro\meter}$ & $>\SI{1.9}{\nano\meter}$
\end{longtable}
}
\endgroup

\end{document}